\documentclass[
superscriptaddress,
preprintnumbers,
amsmath,
amssymb,
prb,
twocolumn,
]{revtex4-2}

\usepackage[left=1.8cm, right=1.8cm, top=1.8cm, bottom=2.0cm]{geometry}
\usepackage[english]{babel}
\usepackage[utf8]{inputenc}
\usepackage{amsthm}
\usepackage{mathtools}
\usepackage{xcolor}
\usepackage[T1]{fontenc}
\usepackage{csquotes}
\usepackage{setspace}
\usepackage{url}
\usepackage{float}
\usepackage{titlesec}
\usepackage{xspace}
\usepackage{tikz}
\titlespacing\subsubsection{0pt}{12pt plus 4pt minus 2pt}{0pt plus 2pt minus 2pt}

\usepackage[
    colorlinks,
    linkcolor = blue,
    citecolor = blue,
    urlcolor = blue
]{hyperref}

\newcommand{\crysviz}{\texttt{CrysViz}\xspace}

\begin{document}

\title[\crysviz: A Portable, Browser-Based Tool for Crystal Structure Visualisation and Analysis]{\crysviz: A Portable, Browser-Based Tool for\\Crystal Structure Visualisation and Analysis}

\author{Abhijith S. Parackal}
\affiliation{Department of Physics, Chemistry and Biology, Linköping University, Linköping, Sweden}

\author{Oscar Bulancea-Lindvall}
\affiliation{Department of Physics, Chemistry and Biology, Linköping University, Linköping, Sweden}
\affiliation{Department of Physics of Complex Systems, E\"otv\"os Loránd University, Budapest, Hungary}
\affiliation{MTA-ELTE Lend\"ulet ``Momentum'' NewQubit Research Group, Budapest, Hungary}

\author{Henricus R.A. ten Eikelder}
\affiliation{Department of Physics, Chemistry and Biology, Linköping University, Linköping, Sweden}
\author{Rickard Armiento}

\affiliation{Department of Physics, Chemistry and Biology, Linköping University, Linköping, Sweden}

\author{Florian Trybel}
\email{florian.trybel@liu.se}
\affiliation{Department of Physics, Chemistry and Biology, Linköping University, Linköping, Sweden}

\begin{abstract}
Visualisation, modification and analysis of crystal and molecular structures is essential in materials science, physics, and chemistry. We present \crysviz (\url{crysviz.org}), a lightweight, open-source\footnote{\url{https://github.com/CrysViz/crysviz}} (AGPL-3.0-or-later) crystal structure visualisation and analysis tool that runs entirely on the user's device in a standard web browser, requiring neither a backend server nor an installation step. Structures are read as CIF files or from input/output files of popular computational software, e.g., VASP and Quantum~ESPRESSO, or retrieved from OPTIMADE-compatible databases. Beyond an extensive set of visualisation features with both fast and high-fidelity modes (including ray tracing and path tracing for publication-quality figures), \crysviz allows symmetry analysis and refinement, distance/angle measurement, bond-length histograms, and the playback of relaxation and molecular dynamics trajectories, which can also be computed on the device with different machine-learned interatomic potentials, all executed locally via WebAssembly. It can also be embedded as a widget in other websites. We describe the design choices that make \crysviz portable across essentially any device and browser, its customisability (per-species colour schemes, user-defined bond cutoffs, selectable rendering pipelines, etc.), and how it complements existing visualisation tools.
\end{abstract}

\maketitle

\section{Introduction}
\begin{figure*}
    \centering
    \includegraphics[width=1\textwidth]{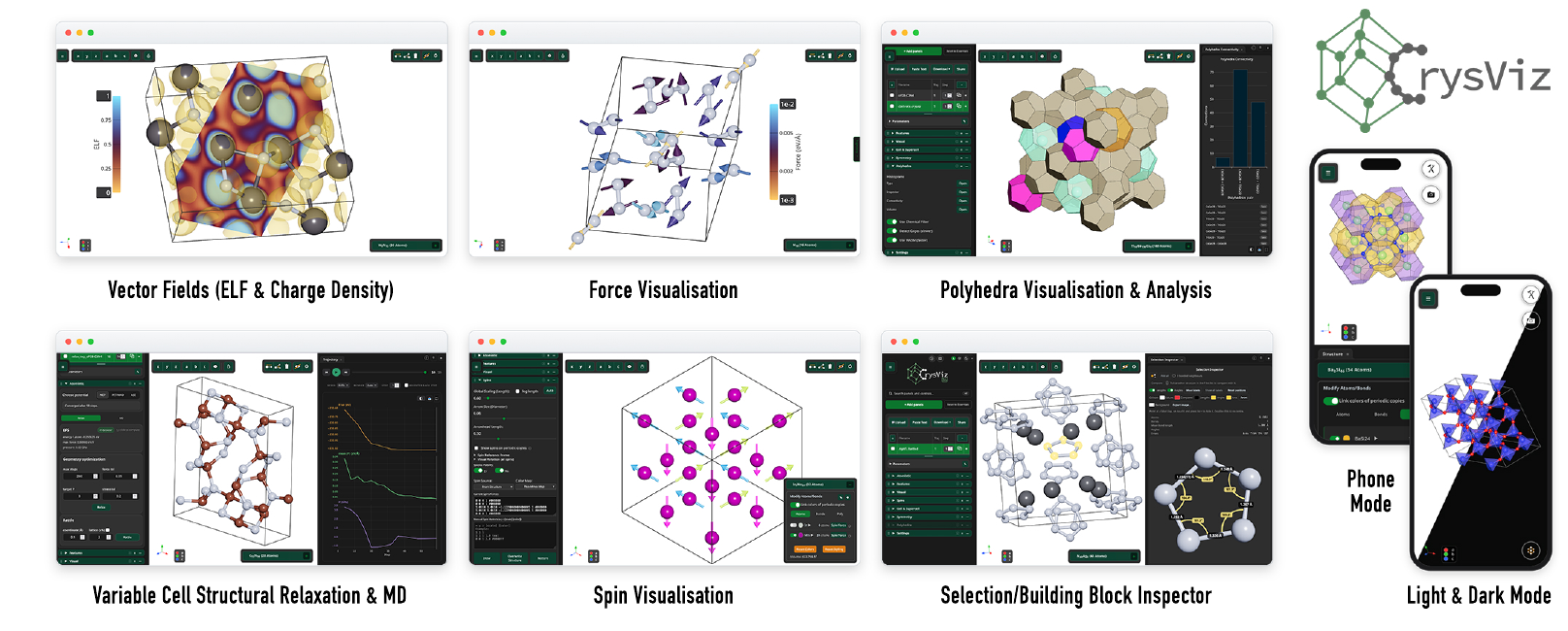}
    \caption{\textbf{Selected examples of features implemented in \crysviz:} Volumetric fields (e.g.\ charge densities or electron localisation functions (ELF)), forces and spins can be directly visualised from calculation outputs. \crysviz supports automatic polyhedra identification, the analysis of variable-cell structural relaxations and molecular dynamics simulations read from DFT or MLIP-based simulation output (extended XYZ, VASP, QE, etc.) or run in \crysviz, and the analysis of a selection of atoms, e.g., rings or molecules, with all distances and angles highlighted. \crysviz features several themes with dark and light layout as well as a dedicated phone mode for small-screen devices.}
    \label{fig:summary}
\end{figure*}
The interactive visualisation of atomic structures is crucial in solid-state physics, materials science and chemistry for gaining both experimental and computational understanding of crystal and molecular structures.  Inspection of experimentally refined or computationally relaxed structures, the analysis of molecular dynamics (MD)
trajectories, or symmetry analysis are only some examples of recurring tasks. There is a range of established desktop tools for this purpose, e.g.,
\textsc{VESTA}~\citep{Momma2011}, \textsc{OVITO}~\citep{Stukowski2010}, \textsc{VMD}~\citep{Humphrey1996},
\textsc{XCrySDen}~\citep{Kokalj1999,Kokalj2003}, \textsc{Mercury}~\citep{Macrae2020}, and
\textsc{Avogadro}~\citep{Hanwell2012}. 

Web-based tools offer a complementary route to these desktop applications: (1) they run on any device with a modern browser, including increasingly powerful tablets and smartphones; (2) new functionality and updates can be deployed instantly to all users; and (3) they require neither installation nor administrator rights, which is convenient on shared machines such as HPC systems or laboratory and beamline computers. For a long time, sophisticated scientific tools have been out of scope for the limited performance of browser-based applications, but recent progress in browser technologies has mostly removed this limitation. Numerically intensive code can now be compiled to WebAssembly (WASM), which runs at a speed approaching that of native code~\citep{Haas2017WASM}, and hardware-accelerated three-dimensional graphics are available through standard browser interfaces, including support for most mobile devices. 

Web technologies now also allow advanced visualisation to be combined with in-browser manipulation and simulation in an integrated, studio-like experience. Recent progress in theoretical models and their efficient implementation enables universal machine-learned interatomic potentials (uMLIPs), trained on large sets of density-functional-theory (DFT) data, to provide useful first approximations for a wide range of material and chemical systems~\citep{Batatia2023Foundation,Mazitov2025PETMAD, liang2026nep89}. Their computational effort is low enough that they can be evaluated on-the-fly in a browser~\citep{mlipcpp}. Interactive visualisation can thus be combined with structural relaxations and short MD simulations of systems with up to a few hundred atoms, which is useful, e.g., for prototyping, during experimental campaigns, or as a visual aid when teaching. With the current rate of improvement of both MLIPs and on-device computational capabilities, production simulations at \textit{ab-initio} accuracy on virtually any device may soon be within reach.

Several web-based tools for molecules and crystal structures are available, from viewing applications such as \textsc{JSmol} (based on Jmol)~\citep{hanson2013jsmol}, \textsc{3Dmol.js}~\citep{3Dmol}, \textsc{NGL}~\citep{NGL}, and \textsc{Mol*}~\citep{sehnal2021mol}, to dataset-centred viewers and frameworks such as \textsc{Crystal Toolkit}~\citep{Horton2023}, \textsc{MatterViz}~\citep{riebesell_matterviz_2022}, and \textsc{chemiscope}~\citep{Fraux2020}. Python libraries such as \textsc{pymatgen}~\citep{Ong2013} and \textsc{ASE}~\citep{Larsen2017} provide extensive programmatic manipulation and analysis of structures. However, there is no widely used application in which structures and trajectories can be visualised, shared, and further analysed with uMLIPs that does not require installation or a Python environment.

We present \crysviz (Fig.~\ref{fig:summary}), a crystal structure visualisation and analysis tool that uses these advances to integrate into one stand-alone application: (a) highly flexible interactive visualisation; (b) a wide range of tools for crystallographic analysis; (c) native reading of the input and output files of the most common simulation software, including complete trajectories; (d) on-the-fly geometry optimisations and MD with uMLIPs on the user's device; and (e) novel ways to share visualisations and complete sessions.

\crysviz is designed around four main ideas:\\

\textbf{(i) Portability}: a single version runs unchanged on any device with a modern browser, from a workstation to a smartphone.\\

\textbf{(ii) On-Device Computation and Data Locality}: Rendering, symmetry analysis, and MLIP-based simulations all run locally on the user's device. Structures are not uploaded to a server unless users explicitly connect to their own calculation backend. This means unpublished or proprietary work can be analysed and full control over the data is maintained. \\

\textbf{(iii) User Expandability}: \crysviz is designed with modularity and expandability in mind. New analysis modules attach to the core viewer through a small add-on interface (Section~\ref{sec:modules}), and the repository contains guides for coding assistants based on large language models.\\

\textbf{(iv) Efficient and Private Data Sharing}: 
The complete session can be saved in a single file, and the current view can be captured in a shareable link or QR code, optionally encrypted with a password (Section~\ref{sec:sharing}). This allows unpublished results to be shared with collaborators even in sensitive surroundings, e.g.\ in industrial collaborations. Links and QR codes can be easily embedded in presentations, posters or articles, and \crysviz can be embedded as an interactive viewer in other web pages (Section~\ref{sec:portability}).\\

\crysviz is free and open-source (Section~\ref{sec:availability}) and depends only on other open-source components. It complements established desktop software by combining zero-installation, cross-device visualisation with computation and analysis performed directly where a structure is displayed.

\begin{figure*}
    \centering
    \includegraphics[width=0.8\linewidth]{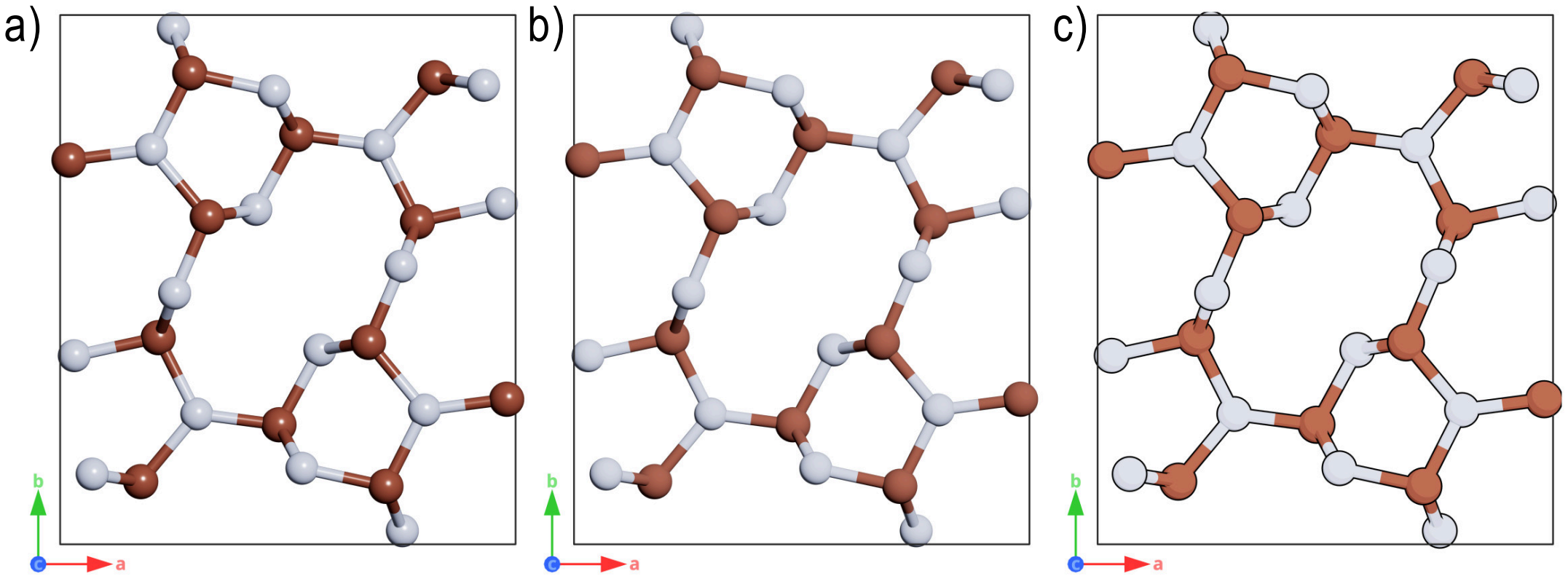}
    \caption{\textbf{Rendering modes:} \textit{oP}28-C$_3$N$_4$ \citep{laniel2025high} with the three different standard rendering styles: \textbf{a)} metallic, \textbf{b)} matte, and \textbf{c)} cel shading.}
    \label{fig:rendering}
\end{figure*}
\section{Design and implementation}
\label{sec:implementation}

\crysviz runs entirely within the web browser: there is no server, database, or user account, and nothing to install (Section~\ref{sec:portability}). The application is a static web page that can be served from any host (in our deployment, GitHub Pages), and all file parsing, rendering, symmetry analysis, and simulation are performed on the user's device.

\begin{figure*}
    \centering
    \includegraphics[height=0.5\linewidth]{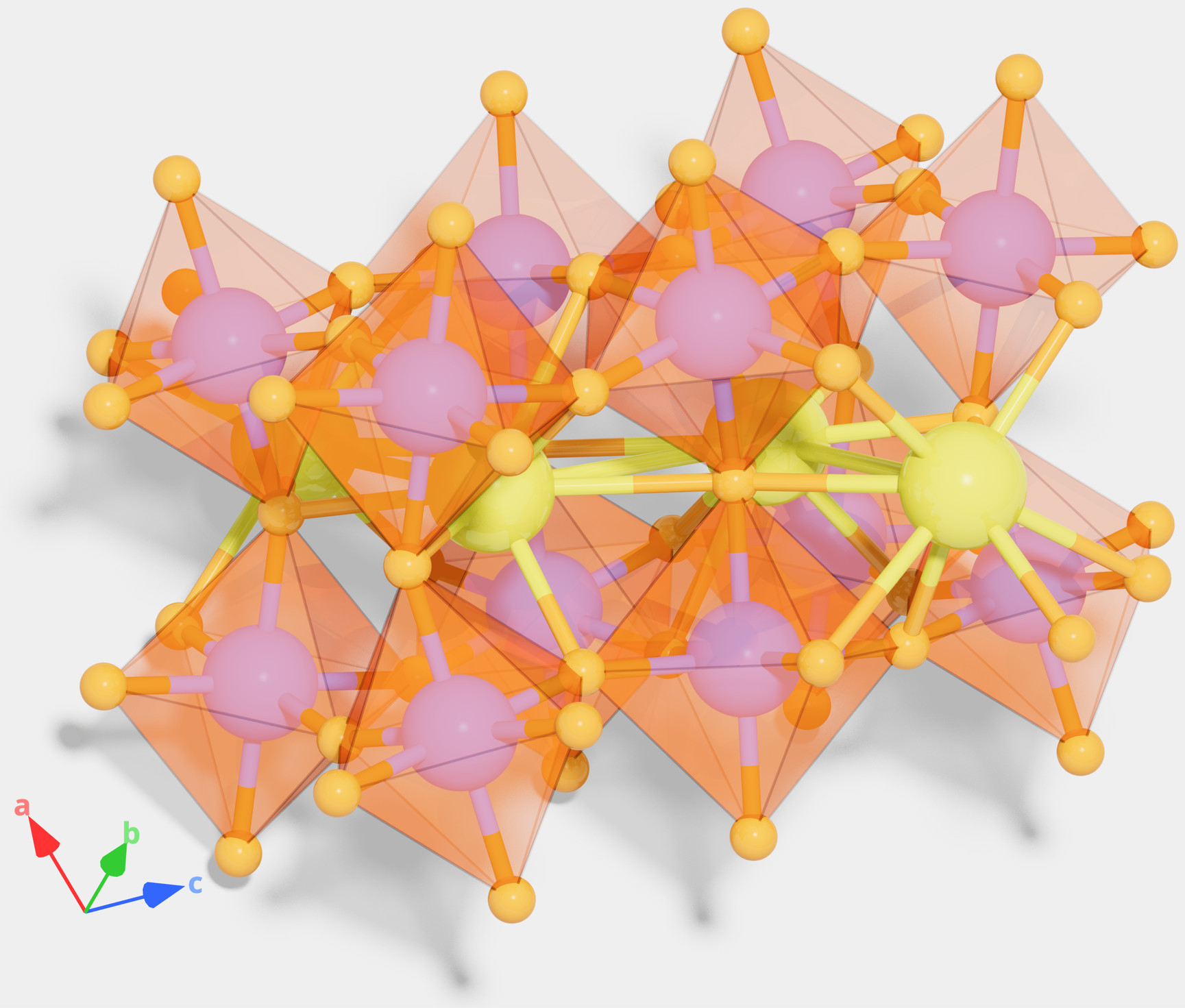}
    \includegraphics[height=0.5\linewidth]{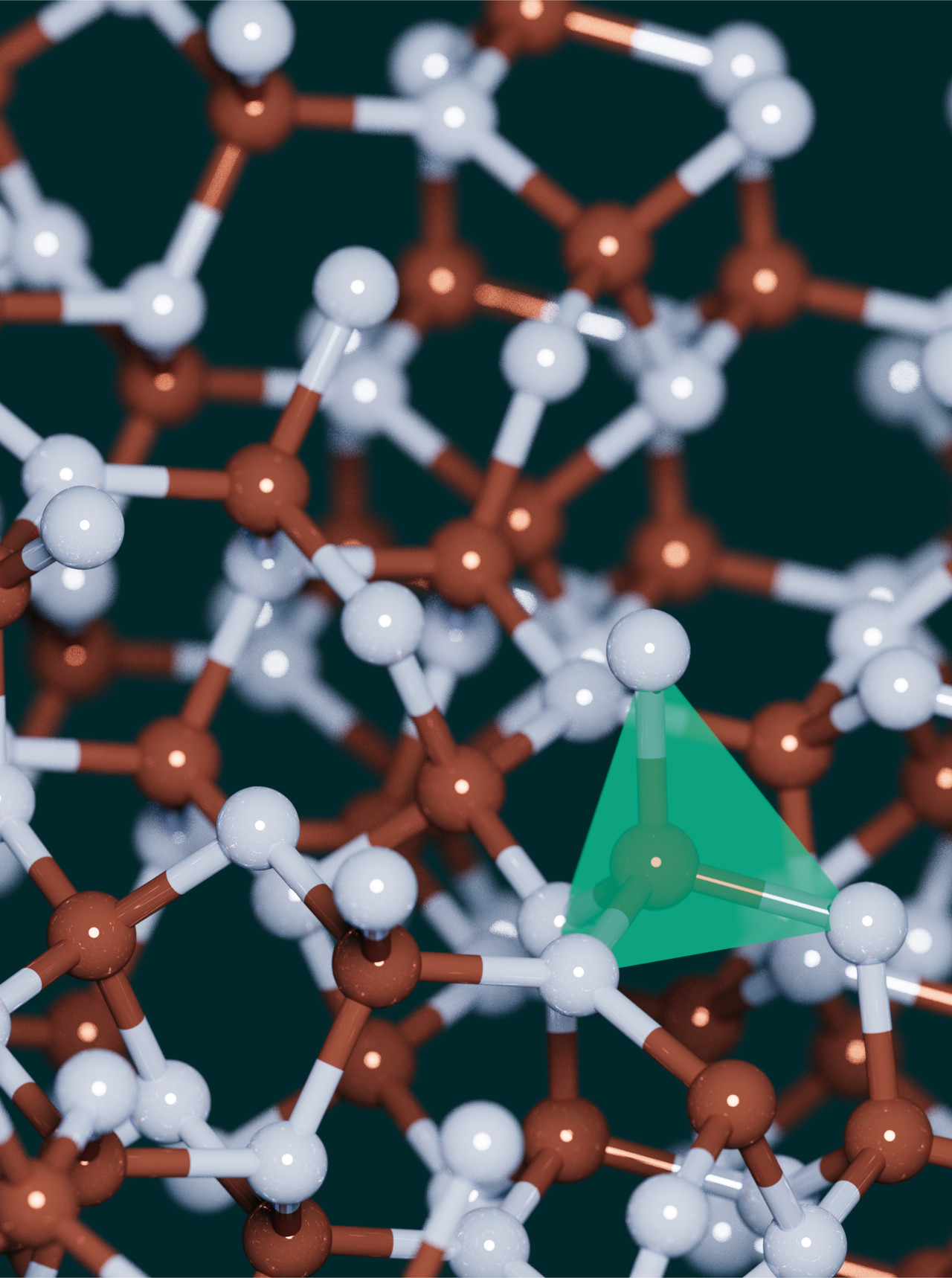}
    \caption{\textbf{Ray and path tracing:} Examples of the ray and path tracing rendering modes. The left figure shows CaTiO$_3$~\citep{ali2005space}  using ray tracing with soft shadows and several layers of semi-transparent polyhedra. The right figure shows \textit{oP}28-C$_3$N$_4$ using path tracing with one tetrahedron highlighted and depth of field aperture used to blur atoms in the background.}
    \label{fig:ray-tracing}
\end{figure*}

\subsection{Rendering}

\crysviz renders structures in interactive 3D using \textsc{three.js}~\citep{ThreeJS}. Because the tool is designed to run on hardware ranging from smartphones to workstations, it offers a choice of rendering modes that
trade rendering speed against image quality. Semi-transparent objects, such as low-opacity atoms or polyhedra, often overlap, and drawing them correctly requires taking their depth order into account. Therefore, \crysviz has a range of rendering engines for depth-aware visualisation of transparent objects, e.g., depth peeling~\citep{Everitt2001,github_depthpeeling} and weighted blended order-independent transparency (WBOIT) \citep{McGuire2013WBOIT,github_wboit} as well as faster alternatives for very low-powered devices.
Depth peeling adjusts the number of layers depending on whether the structure is moving or static, so that even modern smartphones can handle the advanced rendering modes. Several material appearances (shiny, matte, and a flat outlined style) are available (Fig.~\ref{fig:rendering}). For cover-image-quality figures, optional ray-tracing~\citep{Whitted1980,github_raytracing}
and path-tracing \citep{Kajiya1986,github_pathtracing} modes render reflections, refraction, shadows, and light scattering (Fig \ref{fig:ray-tracing_comp}). In these modes, atoms, bonds, and polyhedra can be given, e.g., metallic, glass, or emissive materials, and an optional ground plane allows realistic shadows. Journal-cover-type images (Fig. \ref{fig:cover_style}, \ref{fig:Aperature}, and \ref{fig:Aperature2}) can thus be created within seconds to minutes, without additional rendering software (Fig.~\ref{fig:ray-tracing}). While a structure is rotated, \crysviz automatically switches to depth peeling; when it stands still, the ray- or path-traced image  is re-rendered.

\subsection{On-device computation}

Numerically demanding calculations are compiled to WASM modules. Symmetry detection and refinement use \textsc{Moyo}~\citep{Shinohara2025}, a re-implementation of the widely used
\textsc{spglib}~\citep{Togo2024}. There are custom WASM modules for the detection of bonds and coordination polyhedra, for converting volumetric data to isosurfaces, and for reconstructing wavefunctions from VASP \texttt{WAVECAR} files. %
Built-in structural relaxation and MD use the MLIP engines of Section~\ref{sec:mlip} also via WASM, in some cases making use of GPU-acceleration. Equation-of-state fitting likewise runs on the device, using SciPy~\citep{Virtanen2020SciPy} via
Pyodide~\citep{Pyodide}. Long calculations run in the background so that the interface remains responsive.

\subsection{Input, output, and databases}
\label{sec:io}

\crysviz reads structures and trajectories from VASP~\citep{Kresse1996} (\texttt{POSCAR}, \texttt{CONTCAR}, \texttt{OUTCAR}), Quantum~ESPRESSO~\citep{Giannozzi2009} input and relaxation output, CASTEP~\citep{Clark2005}, FHI-aims~\citep{Blum2009}, SHELX/AIRSS~\citep{Sheldrick2008,Pickard2011} \texttt{.res} files, CIF and magnetic CIF (including fractional occupancies), extended XYZ, and ASE trajectories; volumetric fields from Gaussian \texttt{.cube} and VASP \texttt{CHGCAR}, \texttt{ELFCAR}, and \texttt{WAVECAR} files; and phonon calculations from \textsc{phonopy}~\citep{Togo2023}. TDEP phonon dispersions can also be imported when provided in the phonopy-compatible \texttt{band.yaml} format \citep{knoop2024tdep}. The file type is recognised from its content. Output files are imported as full trajectories with per-frame energies, forces, stresses, and magnetic moments where present. CIF files can be loaded either as \textit{P}1 structures or using the symmetry information contained in the file (Section~\ref{sec:symmetry}). While \crysviz designed in general adapted for periodic crystal structures, it can also visualise and analyse XYZ files with molecular structures (Fig. \ref{fig:structure}). Structures can be loaded, also in bulk, by dragging them into the visualisation area or the file browser (Section~\ref{sec:trajectories}), or with the load structure button.

Structures can further be fetched from any OPTIMADE-compatible provider~\citep{Andersen2021,Evans2024}.
Some providers block cross-origin (CORS) requests from the browser; for providers that allow them, e.g., Alexandria~\citep{Schmidt2023}, \crysviz fetches structures directly by their identifiers. For others, the user can open the OPTIMADE JSON response in a browser tab and paste it into \crysviz, or load it as a file.
Structures and trajectories can be exported as a file (POSCAR, CIF with or without symmetry, CASTEP \texttt{.cell}, FHI-aims, or extended XYZ with all frames; Quantum~ESPRESSO structure cards as text), a self-contained \texttt{.crysviz} session file (Section~\ref{sec:sharing}), an image (PNG or SVG) at a chosen resolution, or a share link or QR code.

\subsection{Adaptive layout}

\crysviz adapts its layout to the screen size of the device, with a dedicated phone mode for small screens. Touch input support is available beyond mirroring mouse
controls.  A small number of advanced panels have to adapt layout on the smallest devices, but the core viewing, measurement, and symmetry workflows are consistently available on all devices.
Different functionalities are organised in panels that can be opened in the left panel container, as floating panels in the structure view or in a side dock.
Each panel has a default setting that optimises usability on large as well as small screens, but the complete interface can be customised by the user. New users start with a slim set of essential panels. Further panels can be added on demand, e.g., via the global search (Section~\ref{sec:search}), or are suggested when the loaded data is compatible with a dedicated analysis feature (e.g.\ a trajectory or a volumetric field).
The panel selection and arrangement are stored in the user's browser data and persist on reload; the arrangement can be reset via a dedicated button, and the selection by deleting the stored browser data.

\section{Portability and Multi-Device Compatibility}
\label{sec:portability}

Because nothing is installed and no server is involved, \crysviz runs identically across operating systems and device types. A single link opens the same view on a tablet, a smartphone, or a restricted teaching/laboratory computer without administrative rights (e.g.\ at a beamline), so that structures received by e-mail can be inspected quickly. The corresponding trade-off is that demanding operations (trajectories ($>$ 20k atoms), high-resolution fields, ray or path tracing, and long MD runs) can briefly slow down on a low-powered device. In return, no possibly unpublished structure is uploaded to a third-party server  as even the external calculation backend has to be run by the user.

\crysviz can also be installed as a Python package (\texttt{pip install crysviz}, available on PyPI). It opens structures from the command line in a desktop window and provides a Python interface, e.g.\ to update a structure from a running simulation or to generate figures from a script.

In widget mode, \crysviz is embedded with a single \texttt{<iframe>} as a compact interactive viewer in other web pages, such as structure databases or online supplementary material (\textit{c.f.} Sec. ~\ref{sec:sharing}). The widget shows only the structure view and a small customisable menu (e.g., primitive vs conventional cell, download of the structure as POSCAR/CIF, or opening the current structure in the full \crysviz window). The embedding page can also control the widget, e.g., to link cameras across structures or to step through a trajectory. Just like the full application, everything runs in the users's browser, so the embedding page can be statically hosted.

\begin{figure}
    \centering
    \includegraphics[width=1\linewidth]{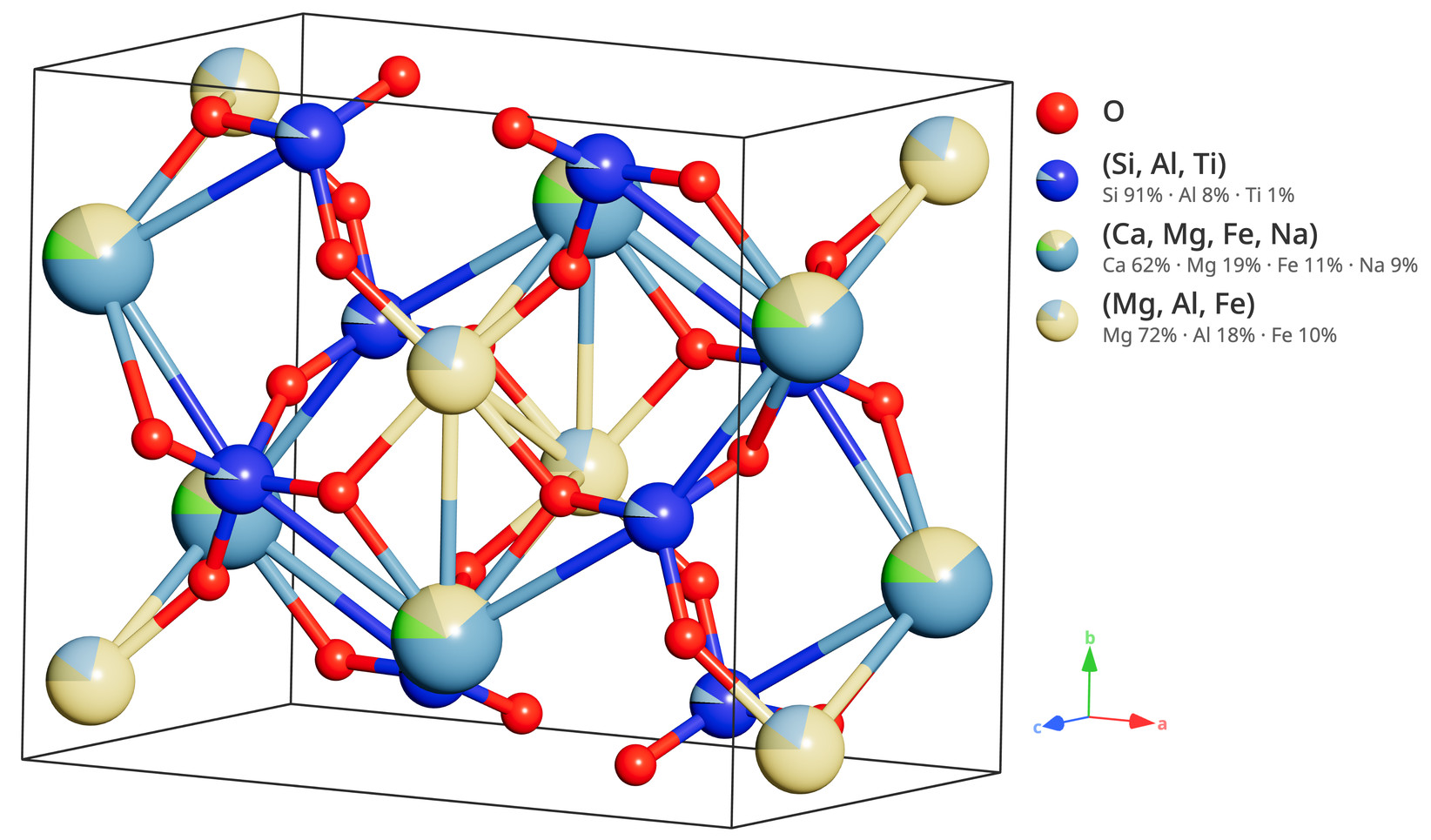}
    \includegraphics[width=1\linewidth]{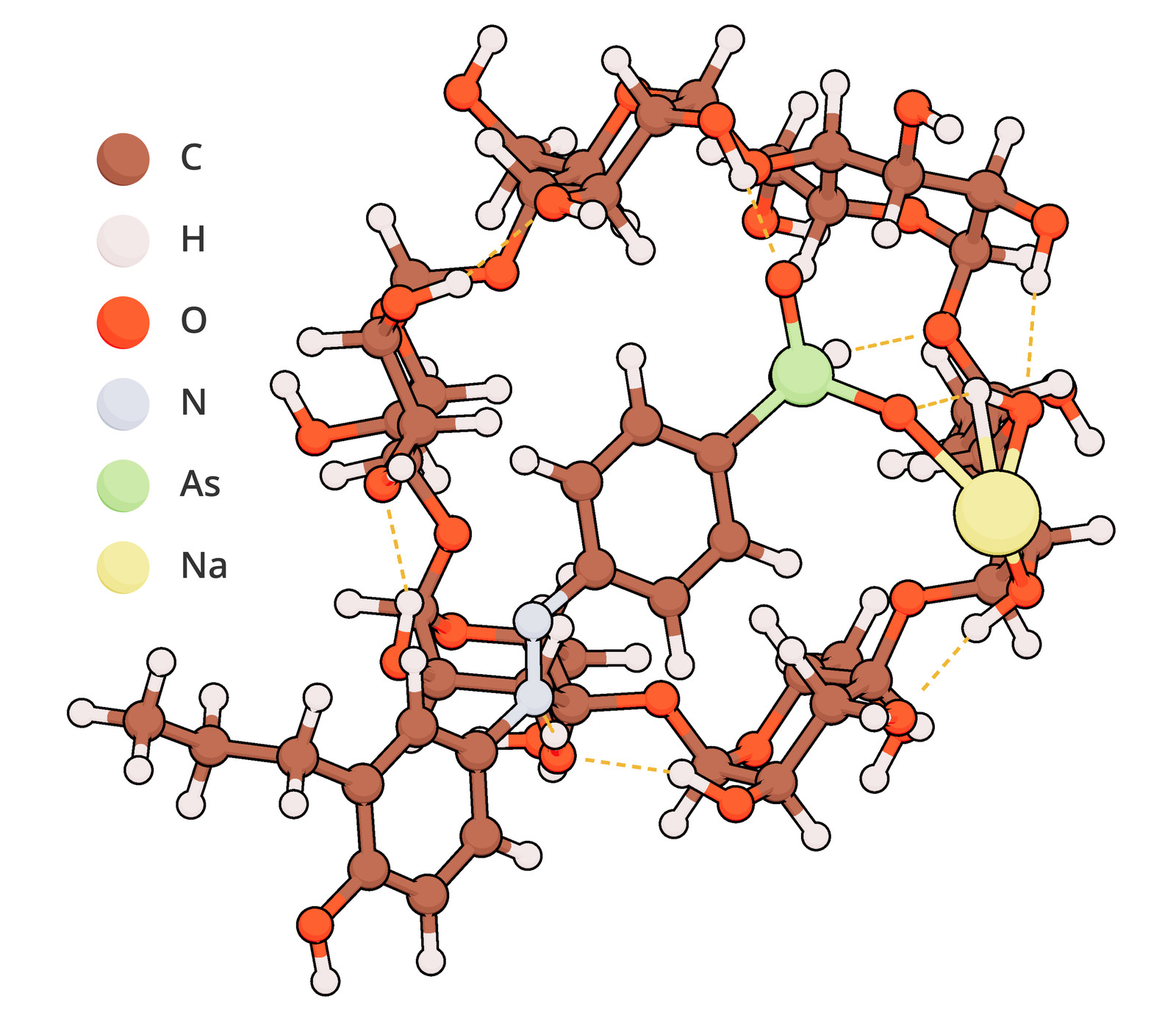}
    \includegraphics[width=0.85\linewidth]{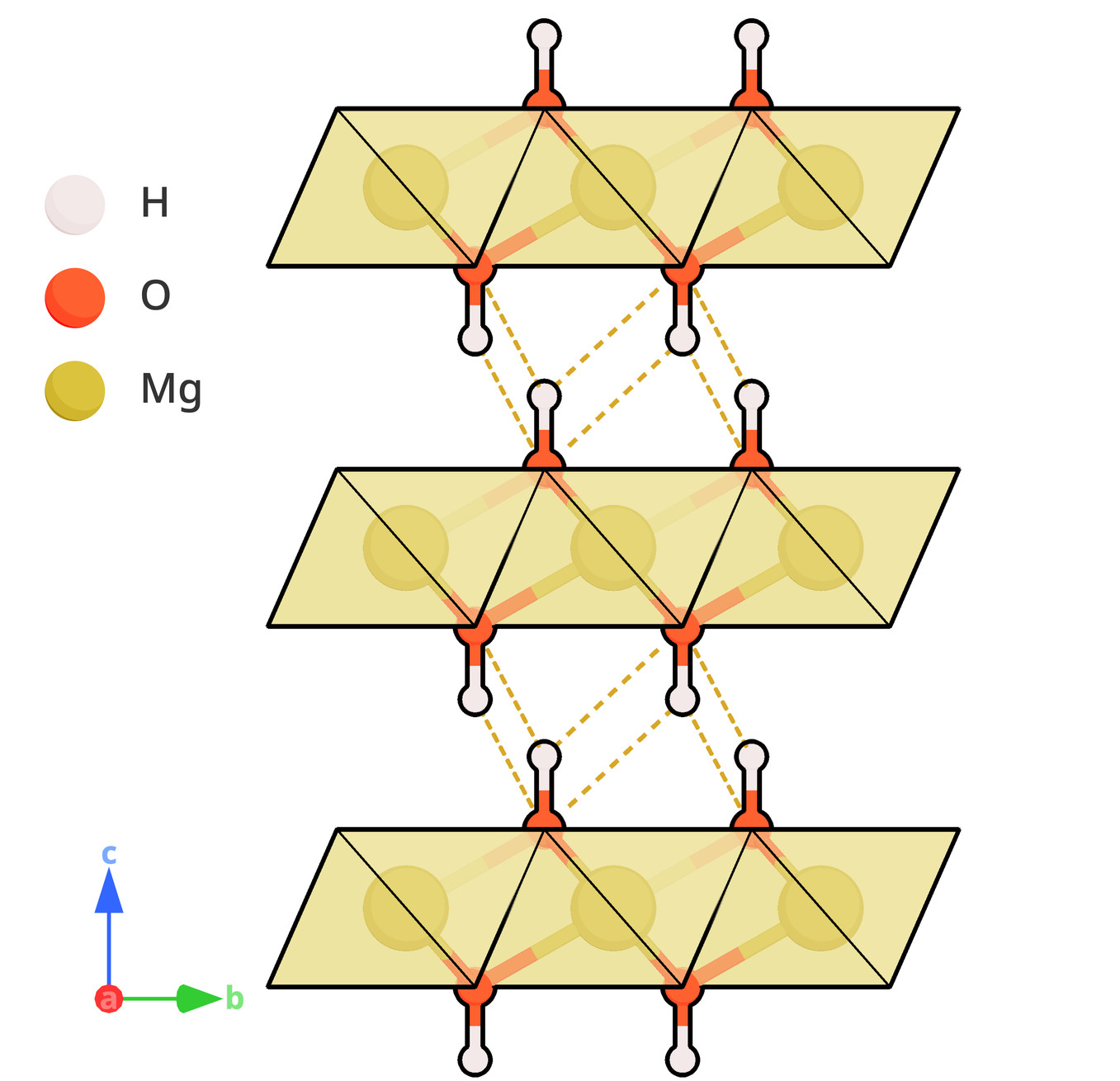}
\caption{\textbf{Partial Occupation, Molecules and Hydrogen Bonds:} \textbf{(top)} The Kakanui augite refined by \citet{Clark1969augite} (COD 1000035), a clinopyroxene with mixed occupation on all three cation sites: Ca/Mg/Fe/Na on M2, Mg/Fe/Al on M1 and Si/Al/Ti on the tetrahedral site. Each site is drawn as a sphere divided into wedges whose sizes give the species' occupancies. \textbf{(middle)} Host--guest complex 12 from the S30L-CI dataset (165 atoms with a Na ion) \cite{sure2015s30l} and \textbf{(bottom)} brucite, Mg(OH)$_2$ \citep{chakoumakos1997low}. Dashed lines show hydrogen bonds in the lower two visualisations.}
\label{fig:structure}
\end{figure}

\section{Visualisation and analysis features}
\label{sec:features}

\crysviz provides a broad set of visualisation and analysis features, of which some are summarised below. All features are documented in full in the user manual~\citep{crysvizmanual}.
\subsection{Global Search}
\label{sec:search}
\crysviz features a global search for all functionalities and nearly all individual settings, as well as atoms, bond pairs, and Wyckoff sites of the loaded structure. It overloads the standard search shortcut (\texttt{Ctrl+F}) and has a search bar in the top left corner under the logo. Names are fuzzy-matched, tolerating spelling errors. Furthermore, typing for example 222 creates a $2\times2\times2$ supercell (up to $9\times9\times9$). The name of a feature followed by \blockquote{on} or \blockquote{off} allows toggling features, e.g., \blockquote{Bonds on} or \blockquote{Bonds off} switches on/off bond visualisation. Panels for functionalities that are not shown in the current selection can be searched and added by confirming with enter. A growing catalogue of phrases is matched to functionalities, e.g., \blockquote{lattice vibrations} opens the phonon visualisation.

\subsection{Structure Display} Atoms, bonds, and coordination polyhedra are drawn with per-species and per-atom control over colour, size, and opacity. Bonds follow user-adjustable element-pair cutoffs. Bonded neighbours, parts of polyhedra and rings (3- to 8-membered) lying just outside the drawn cell can be automatically detected and completed, so that bonds or building blocks are not left incomplete at the cell boundary. Hydrogen bonds (D--H$\cdots$A) are detected automatically from the H$\cdots$A distance and the D--H$\cdots$A angle and drawn as dashed lines~\citep{Steiner2002} (Fig.~\ref{fig:structure}). Their distance range, colour, and visibility can be set for each hydrogen--acceptor element pair. Atoms and bonds can also be coloured according to a numerical quantity (Section~\ref{sec:custom}).

Periodic images are shown by default, supercells can be created on the fly and vacuum added to a structure. A general lattice transformation ($3\times4$ matrix) allows reshaping the cell while preserving styling. Fractional occupations and charges are read from CIF files, and an \emph{Order Structure} tool allows constructing ordered approximations when a definite species per site is required (Section~\ref{sec:mlip-runs}). Partially occupied sites are drawn as spheres divided according to their occupancies (Fig.~\ref{fig:structure}) with automatically generated legends (user editable). Formal charges can be shown next to the atoms are read either from a CIF file or can be set by the user. All UI elements shown in the structure view, e.g., axis compass, colormaps, and legends, can be freely resized and placed by the user.

\begin{figure}
    \centering
    \includegraphics[width=\columnwidth]{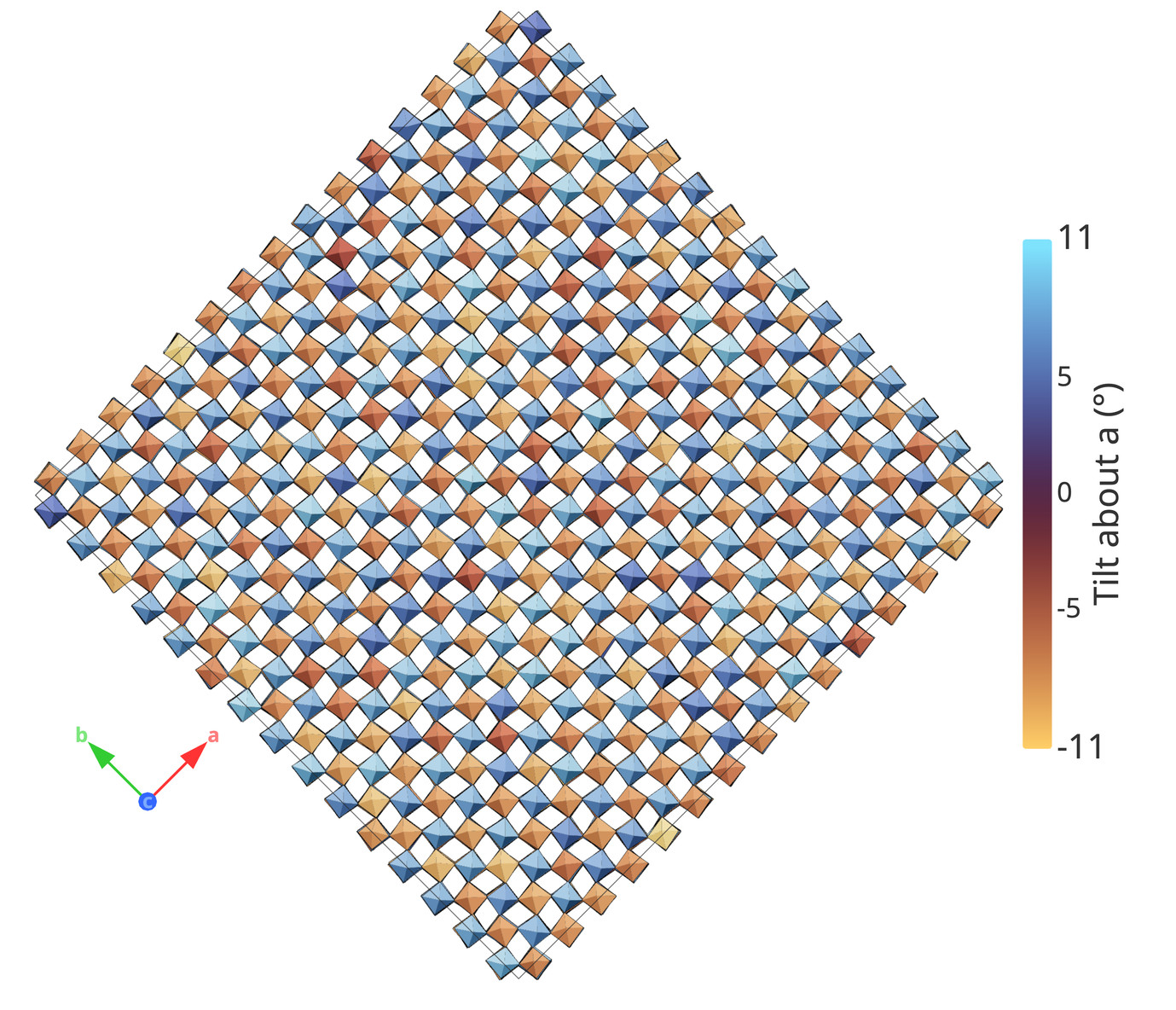}\\[0.5em]
    \includegraphics[width=\columnwidth]{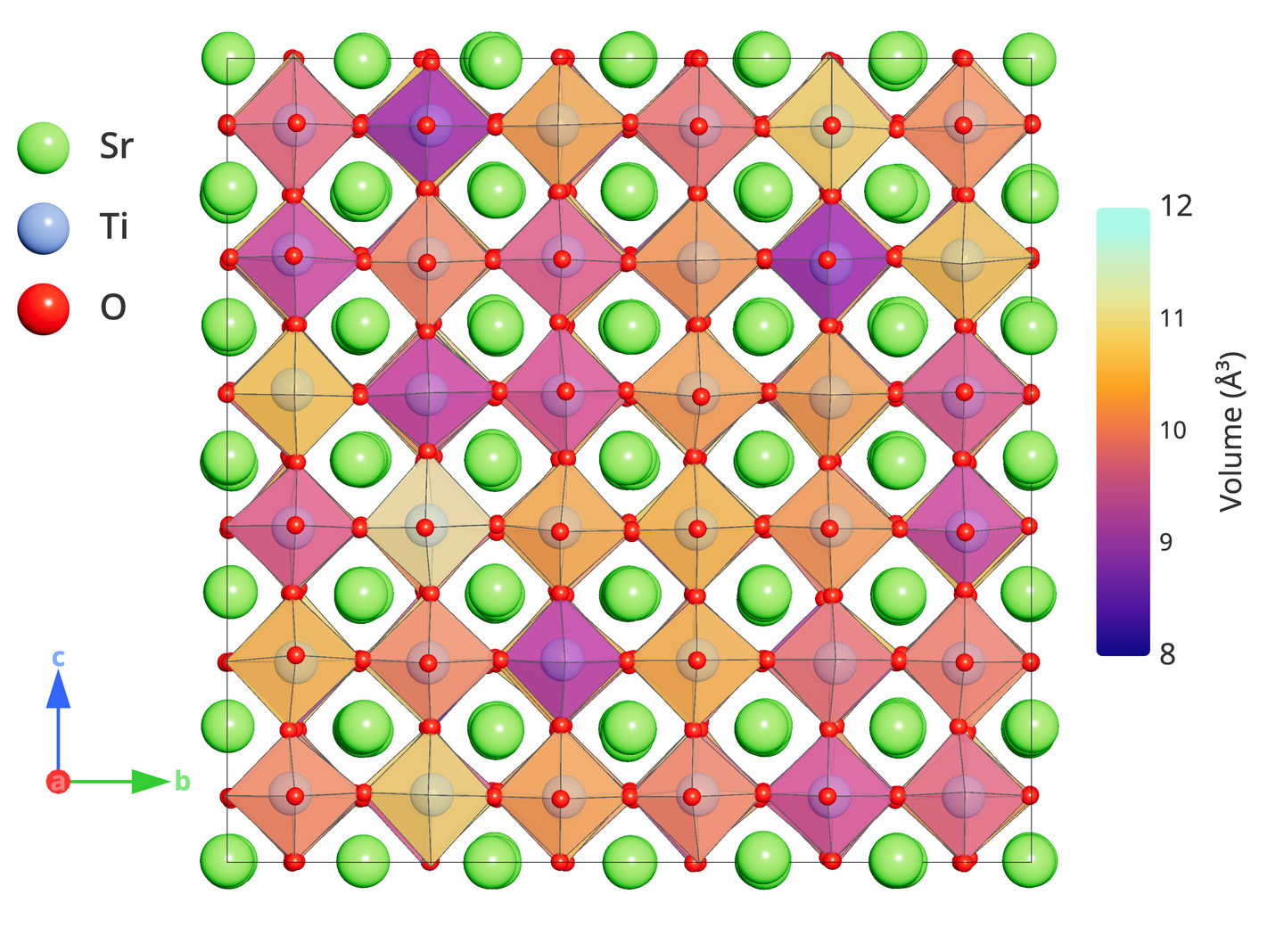}
\caption{\textbf{Polyhedra Visualisation:}
\textbf{(top)} Snapshot from $15\times15\times8$ supercell trajectory of CaTiO$_3$ with 36\,000 atoms viewed along $c$, atoms hidden, octahedra coloured by their tilt about $a$ against the automatically found pseudo-cubic axes.\textbf{(bottom)} $6\times6\times6$ supercell of cubic SrTiO$_3$ viewed along $a$, octahedra coloured by their volume. In both cases the images are created from artificial data with thermal noise is added.}
\label{fig:polyhedra}
\end{figure}

\subsection{Building and Editing Structures} Structures can be built in the browser atom by atom or from a space group and its occupied Wyckoff positions, and loaded structures can be edited live.

\subsection{Symmetry}
\label{sec:symmetry}
The symmetry of a crystal structure can be analysed with \textsc{Moyo}~\citep{Shinohara2025} at a user-adjustable tolerance to determine the space group, reporting the Hermann--Mauguin symbol and number and an AFLOW-style protostructure label. \citep{Mehl2017AFLOW,goodall2022rapid} A structure can be symmetrised in primitive or conventional setting. The real-space asymmetric unit of the space group, i.e.\ the smallest part of the cell from which the symmetry operations generate the full structure, can be displayed as a semi-transparent polyhedron in the conventional cell, using the asymmetric units tabulated in the \textsc{httk} space-group data~\citep{httkdata}, together with its defining inequalities and its volume as a fraction of the cell. Its colour and opacity can be adjusted, and the atoms inside it can be highlighted (Fig.~\ref{fig:asu}). A Wyckoff editing mode groups atoms into Wyckoff orbits and restricts edits to the degrees of freedom permitted by the site symmetry, which allows, e.g., symmetry-restricted structural relaxations and MD runs (Section~\ref{sec:mlip-runs}). When a CIF file is loaded with symmetry information, the full cell is reconstructed from the asymmetric unit and the symmetry operations, and the structure opens in the Wyckoff editing mode, preserving the setting of the structure specified in the file. \crysviz automatically determines the space group of the reconstructed structure and compares it with the one declared in the file. If the coordinates are given with too few digits (e.g.\ 0.333 instead of 1/3), the structure can be symmetrised to the detected space group, which places the atoms exactly on their special positions. \crysviz can display the symmetry elements of a structure's space group directly in the structure view (mirror and glide planes, rotation and screw axes, inversion centres and rotoinversion axes) within the conventional unit cell, using the graphical symbols of symmetry elements of the International Tables for Crystallography, Vol. A,\cite{aroyo2016international} adapted to three dimensions (Fig. \ref{fig:asu}).

\subsection{Measurements and Hiding Atoms} Distances (between two atoms) and angles (among three atoms, with the vertex selected second) can be measured by selecting atoms in the 3D view (Fig.~\ref{fig:Measurements} and \ref{fig:Measurements2}). Measurements update when a trajectory is played and are stored in the session file and the shared link (Section~\ref{sec:sharing}). Atoms can be temporarily hidden and restored without altering the underlying data.

\begin{figure}
    \centering
    \includegraphics[width=1\linewidth]{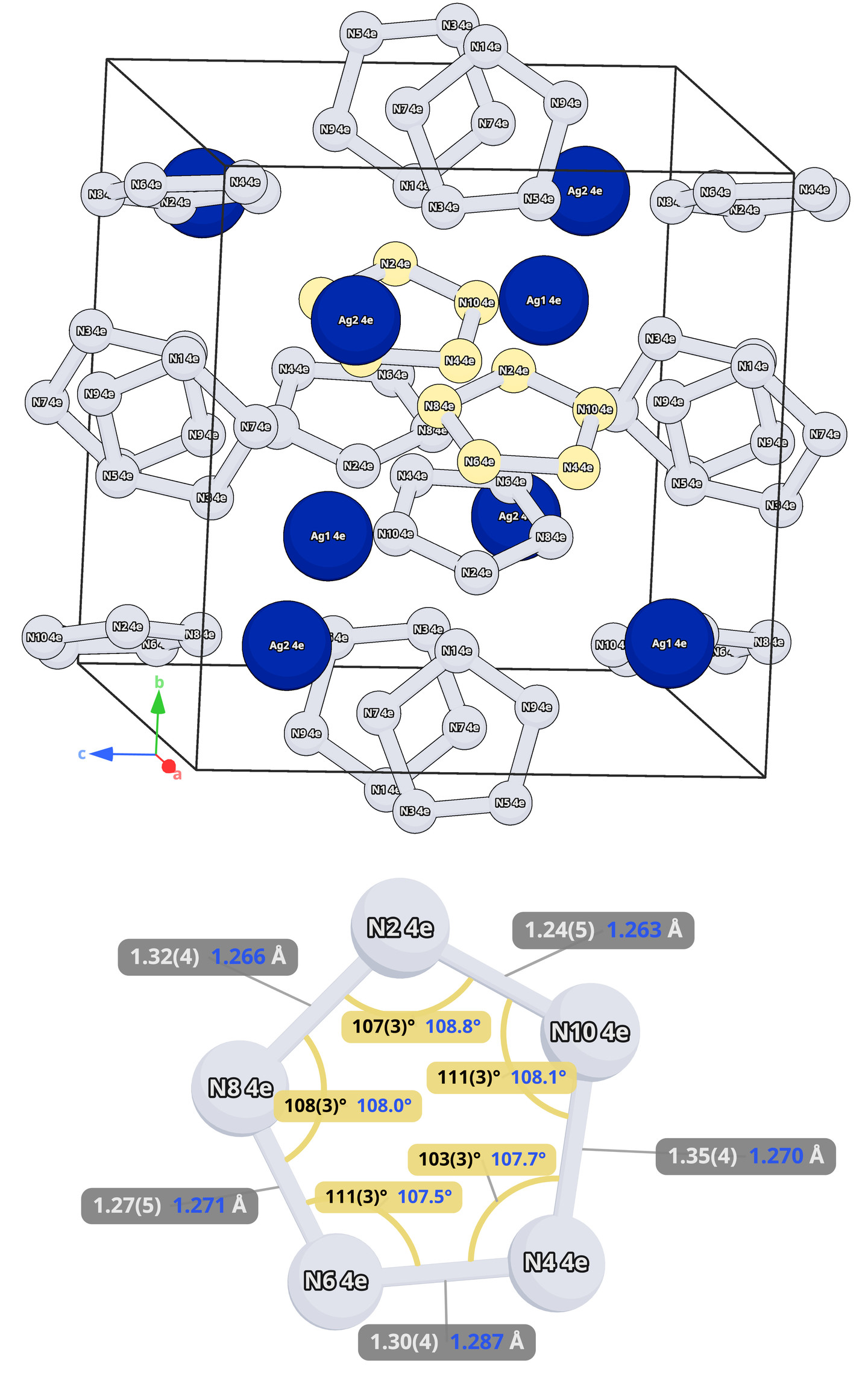}
    \caption{\textbf{Selection Inspector:} Illustration of the analysis of the N$_5$ ring in AgN$_5$ at 118 GPa~\citep{liang2026high}. \textbf{(top)}
   Structure in the cel shading rendering mode, with the atoms labelled by their Wyckoff positions (read from the CIF or determined with \textsc{Moyo}~\citep{Shinohara2025}). Rings cut by the cell boundary can be completed automatically. Highlighted atoms are selected for analysis in the inspector. \textbf{(bottom)} Figure exported from the inspector: The experimental distances and angles are shown in white (including errors propagated from the X-ray diffraction data; CCDC Number: 2475575~\citep{Liang2025_CSD2475575}) in comparison with the result of a variable-cell structural relaxation with the PET-MAD S potential (blue) performed directly in \crysviz. All colours, label positions and sizes can be modified by the user.
}
    \label{fig:inspector}
\end{figure}

\subsection{Polyhedra Analysis} Coordination polyhedra are identified from the bonding cutoffs using a Voronoi-based neighbour criterion, including uncentred cages (for example the icosahedral B$_{12}$ unit). Polyhedra can be analysed by type, by connectivity (corner-, edge-, and face-sharing, together with two bond-bridge categories), and by volume. They can be viewed individually through an inspector (Fig. \ref{fig:tetrahedra_inspector}) that calculates the coordination number, the volume, the bond-length distortion \citep{Baur1974}, and, for tetrahedra and octahedra, the bond-angle variance \citep{Robinson1971}. For large cells ($>$ 2000 atoms) a \textit{fast} mode is automatically activated (there are options for \textit{auto}, \textit{fast}, and \textit{advanced}) employing a faster polyhedra search relying only on the convex hull of each atom's bonded neighbours (no cages are detected). This allows the analysis of trajectories beyond 100k atoms (Fig.~\ref{fig:polyhedra}).

\subsection{Selection Inspector}

\crysviz features a selection inspector that allows to isolate and inspect a sub-selection of the structure (Fig. \ref{fig:inspector}). It allows to show all bond distances and angles with user adaptable labels. If a second structure is selected in the file browser the interatomic distances and angles can be overlayed to compare, e.g., experimental and calculated structures. A separate figure export allows to obtain images as shown in the lower panel of Figure \ref{fig:inspector}.  The inspector can also be used to zoom into a region in a large structure (Fig. \ref{fig:largeMD}).

\subsection{Bonding Analysis} Bond-length and coordination-number histograms  (Fig. \ref{fig:bond_length_coordination}) allow quick analysis of bonding environments. Histograms update live as bond cutoffs are changed or when a trajectory is played; selecting a histogram bin highlights the corresponding bonds or atoms in the 3D view. Enabling Wyckoff modes allows grouping per Wyckoff position (\ref{fig:bond_length_coordination_wyk}).

\subsection{Fields and Planes} 
\begin{figure}
\centering
\begin{minipage}{0.49\linewidth}
\centering
    \tikz{
        \node at (0,0) {\includegraphics[width=\linewidth]{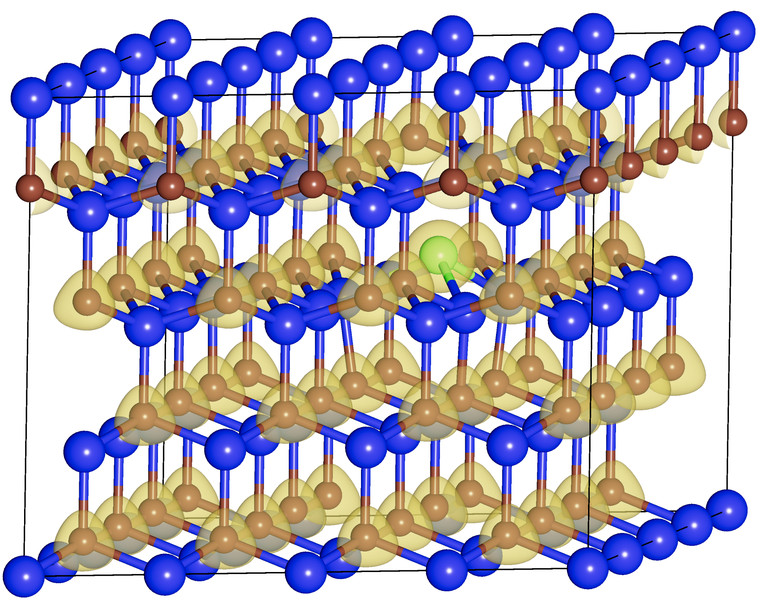}};
        \node at (-2,2) {a)};
    }
\end{minipage}
\hfill
\begin{minipage}{0.49\linewidth}
    \centering
    \tikz{
        \node at (0,0) {\includegraphics[width=\linewidth]{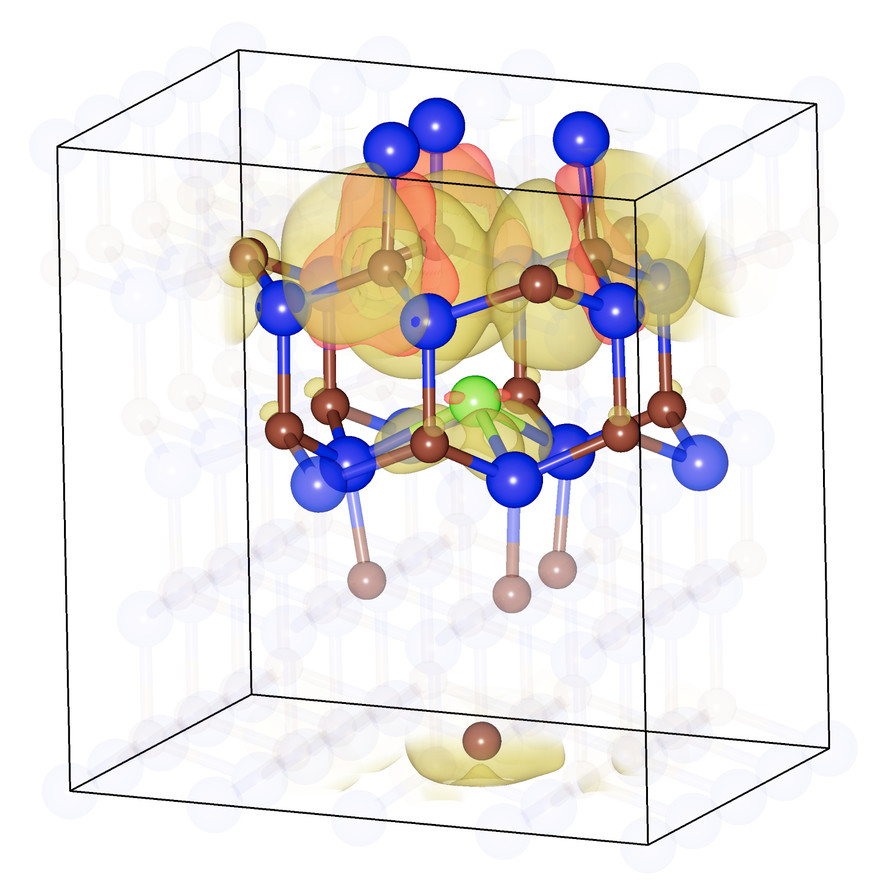}};
        \node at (-2,2) {b)};
    }
\end{minipage}

\begin{minipage}{0.49\linewidth}
\centering
    \tikz{
        \node at (0,0) {\includegraphics[width=\linewidth]{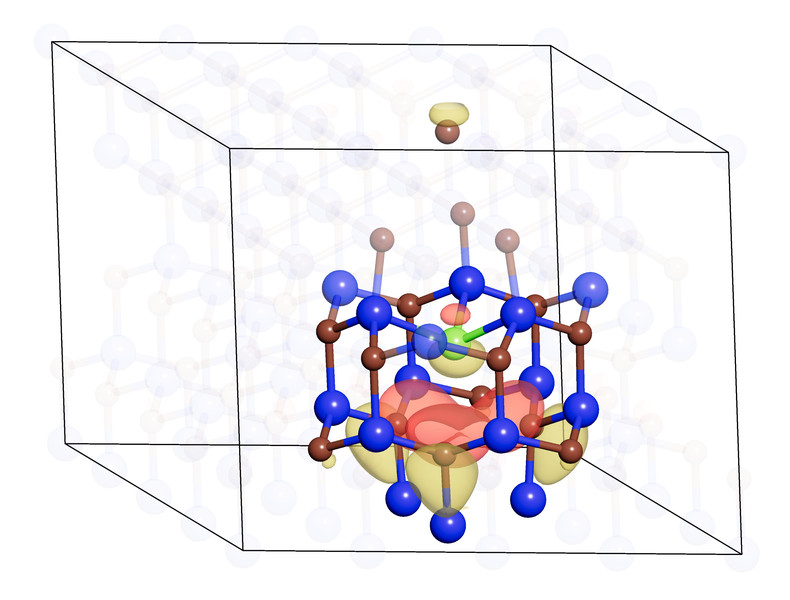}};
        \node at (-2,2) {c)};
    }
\end{minipage}
\hfill
\begin{minipage}{0.49\linewidth}
    \centering
    \tikz{
        \node at (0,0) {\includegraphics[width=\linewidth]{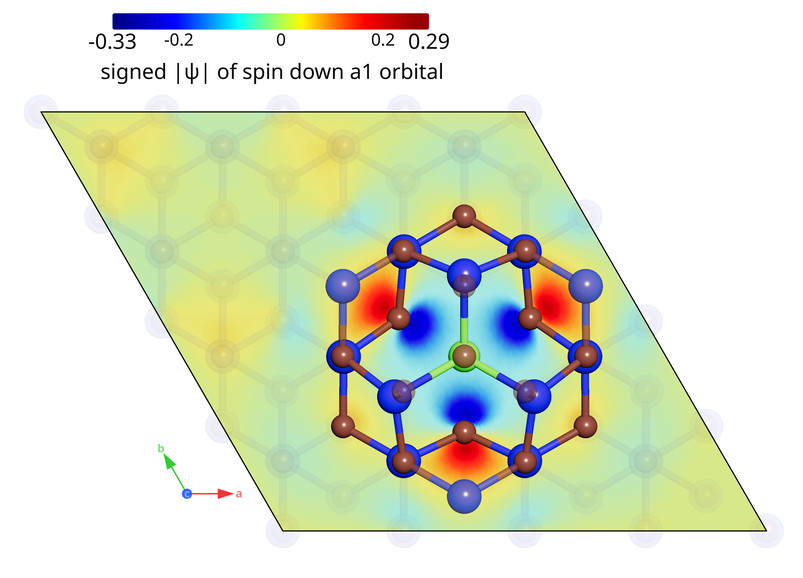}};
        \node at (-2,2) {d)};
    }
\end{minipage}

\centering
\begin{minipage}{0.48\linewidth}
    \centering
    \tikz{
        \node at (0,0) {\includegraphics[width=\linewidth]{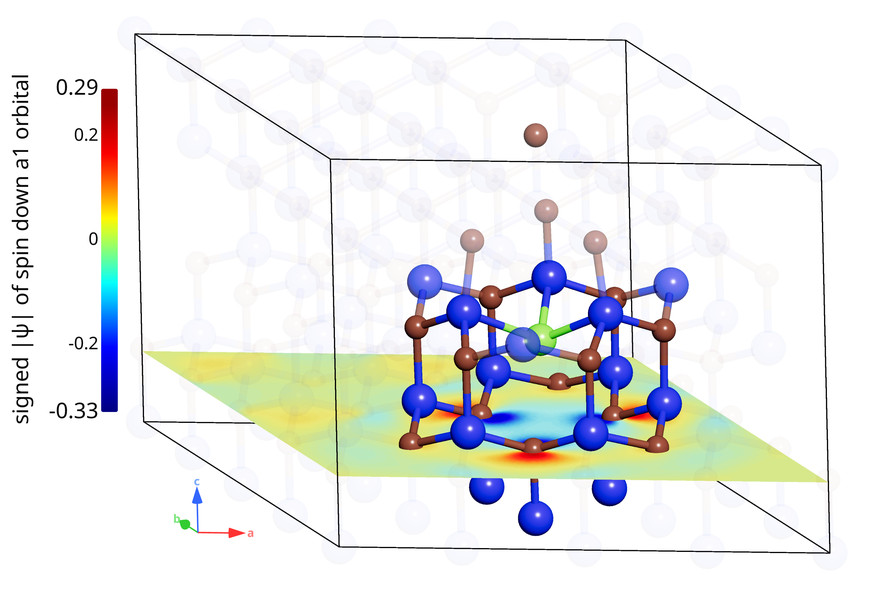}};
        \node at (-2,1.5) {e)};
    }
\end{minipage}

\caption[Volumetric fields]{ \textbf{Volumetric Fields and Wave Functions:} Volumetric data of a chlorine-vacancy center defect in 4H-SiC \cite{bulancea-lindvall_ClV} visualised as field and planar map. \textbf{(a)} Charge-density isosurface as extracted from a standard VASP \texttt{CHGCAR}.
\textbf{(b)} Magnetisation (spin) density with positive (yellow) and negative (red) lobes in different colours. A focus region is enabled to fade anything outside a radial selection, as to focus on visualising the defect vicinity.
\textbf{(c)} A single band from a \texttt{WAVECAR} at a specified k-point and spin channel, shown as $|\psi|$ weighted with the sign of the real-valued part of $\psi$, with focus region. \textbf{(d)} A plane defined as cutting the three (highlighted) nearest-neighbour carbon sites to the defect. The cross section of the band wavefunction signed $|\psi|$ is visualised on the plane surface.
\textbf{(e)} Figure (d) from a different angle, showing the planar cut and position of the plane in the cell, together with the focus region.}
\label{fig:fields}
\end{figure}
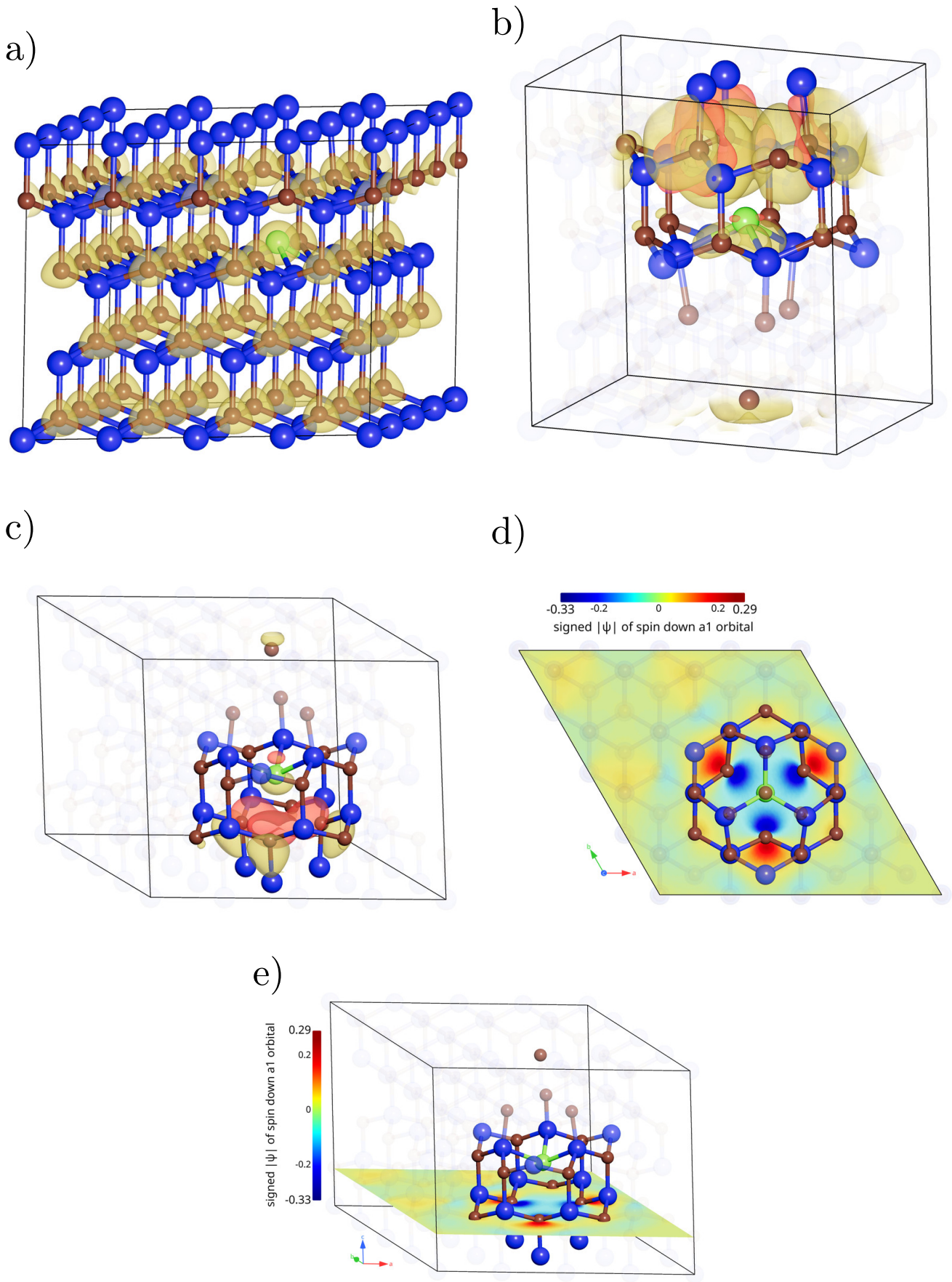

Charge density and the electron localisation function (ELF), including spin-resolved channels, can be viewed via isosurfaces with independent positive/negative colouring and an adjustable (linear or logarithmic) threshold. Wavefunctions of individual bands can be read from VASP \texttt{WAVECAR} files, including non-collinear calculations, and new fields, e.g.\ charge-density differences, can be built as weighted sums of loaded fields. Cut planes may also be defined and used to clip the structure to reveal its interior and to plot field cross-sections. Planes can be created and modified via Miller indices, Cartesian parameters, or automatically created to minimise the root-mean-square distance with respect to a set of $N$ selected atoms  (Fig.~\ref{fig:fields} and Fig. \ref{fig:export}).

\subsection{Focus Regions} In large structures, a region of interest such as a molecule or a defect can be highlighted by fading everything outside a sphere around selected atoms or a chosen coordinate (Fig.~\ref{fig:fields} b, c, and e). The focus mode acts on atoms, bonds, polyhedra, volumetric fields and wave functions.

\subsection{Spins and Forces} 
\begin{figure*}[t]
    \centering
    \includegraphics[width=1\linewidth]{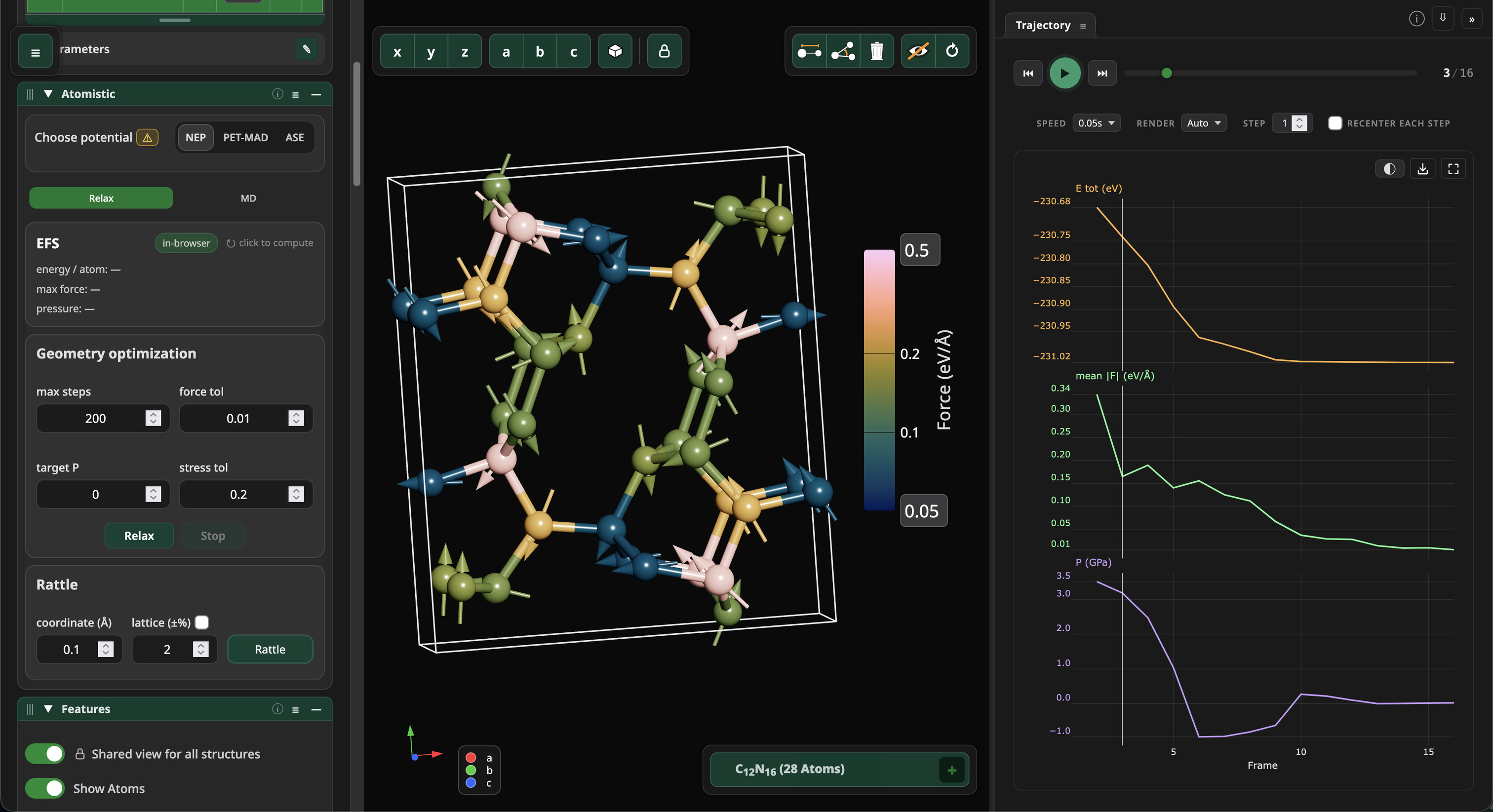}
    \caption{\textbf{Visualisation of Trajectories:} Trajectories from (variable-cell) structural relaxations and MD simulations, read from file (Section~\ref{sec:io}) or computed with the built-in MLIPs, are shown with an interactive plot (Section~\ref{sec:trajectories}). Here, a relaxation computed in \crysviz is shown with the forces as arrows and atom colours. A snapshot can be exported including the colour bar (Fig.~\ref{fig:export}).}
    \label{fig:trajectory}
\end{figure*}

Per-atom forces and magnetic moments (spins) are directly read from simulation output (Section~\ref{sec:io}). In addition, forces can be computed with the built-in MLIPs, and spins can be entered manually. Both are drawn as arrows whose length, thickness, and visibility per species can be adjusted, optionally with a logarithmic length scale. The arrows are coloured by direction or magnitude (Section~\ref{sec:custom}; Fig.~\ref{fig:trajectory}). Magnetic moments that a code reports in a spin-quantisation frame (e.g.\ for a non-default \texttt{SAXIS} in VASP) are rotated into Cartesian coordinates automatically or orientated following a user input.

\subsection{Trajectories}
\label{sec:trajectories}
Multi-frame trajectories can be played back with an adjustable playback speed. Forces, spins, measurements and colour maps are updated frame by frame. The total energy, the mean force, the pressure and, for MD runs in \crysviz, the temperature and its target value are plotted in the trajectory player (Fig.~\ref{fig:trajectory}). The file browser allows switching between several structures and trajectories, combining several structures into a trajectory, and extracting parts or single steps from trajectories.

\begin{figure*}
    \centering
    \includegraphics[width=0.9\linewidth]{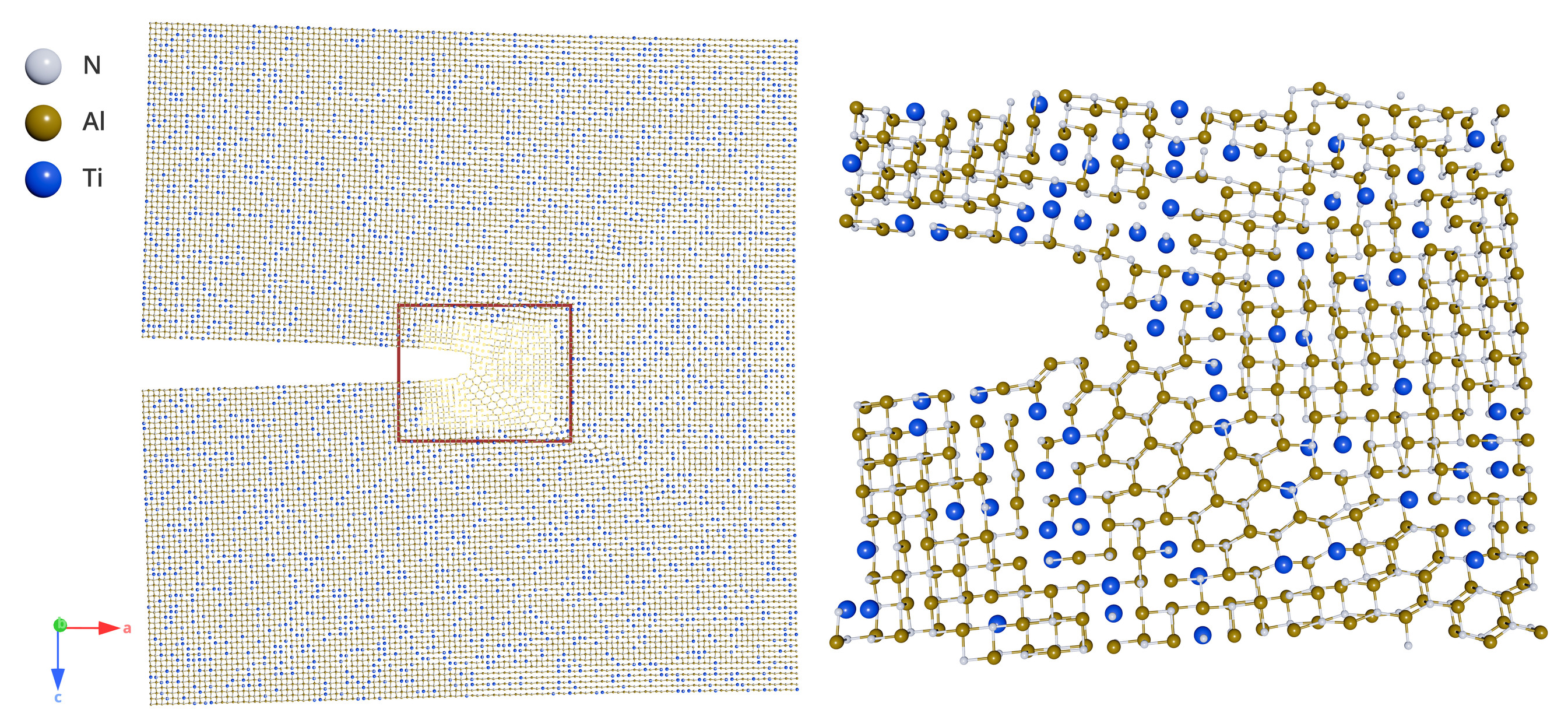}
    \caption{\textbf{Large Scale Structures:} Visualisation of a stress-activated transformation from B$_1$ to B$_{\rm k}$ near the tip in Ti$_{0.2}$Al$_{0.8}$N \cite{sangiovanni2025controlled}. The slice through the simulation box contains 28438 atoms. The building block inspector allows to zoom in on the region close to the crack tip (indicted via the red square) highlighting the relevant area ($\sim$ 1000 atoms) where the material undergoes a phase transformation (hexagonal atomic arrangement).}
    \label{fig:largeMD}
\end{figure*}

Trajectories with several hundred atoms and several thousand time steps can be viewed even on mobile devices, and with several ten thousand atoms on more powerful devices (e.g.\ a MacBook Pro with an M3 Max processor). In the current implementation, trajectories  above 50k atoms with custom colours (Fig. \ref{fig:largeMD}), colour maps and on-the-fly bond calculation start to play at lower frame rates. The currently tested limit lays at about 400k atoms if only atoms and polyhedra are displayed with standard settings in the fast rendering and polyhedra mode. All analysis functionalities show reasonable, strongly device-dependent performance up to about 1000 atoms and trajectories of the order of 5000 steps. Automatic cage identification and completion is the most expensive operation and can require several seconds for large cells.

\subsection{Comparison and Overlay} 
\begin{figure}
    \centering
    \includegraphics[width=1\linewidth]{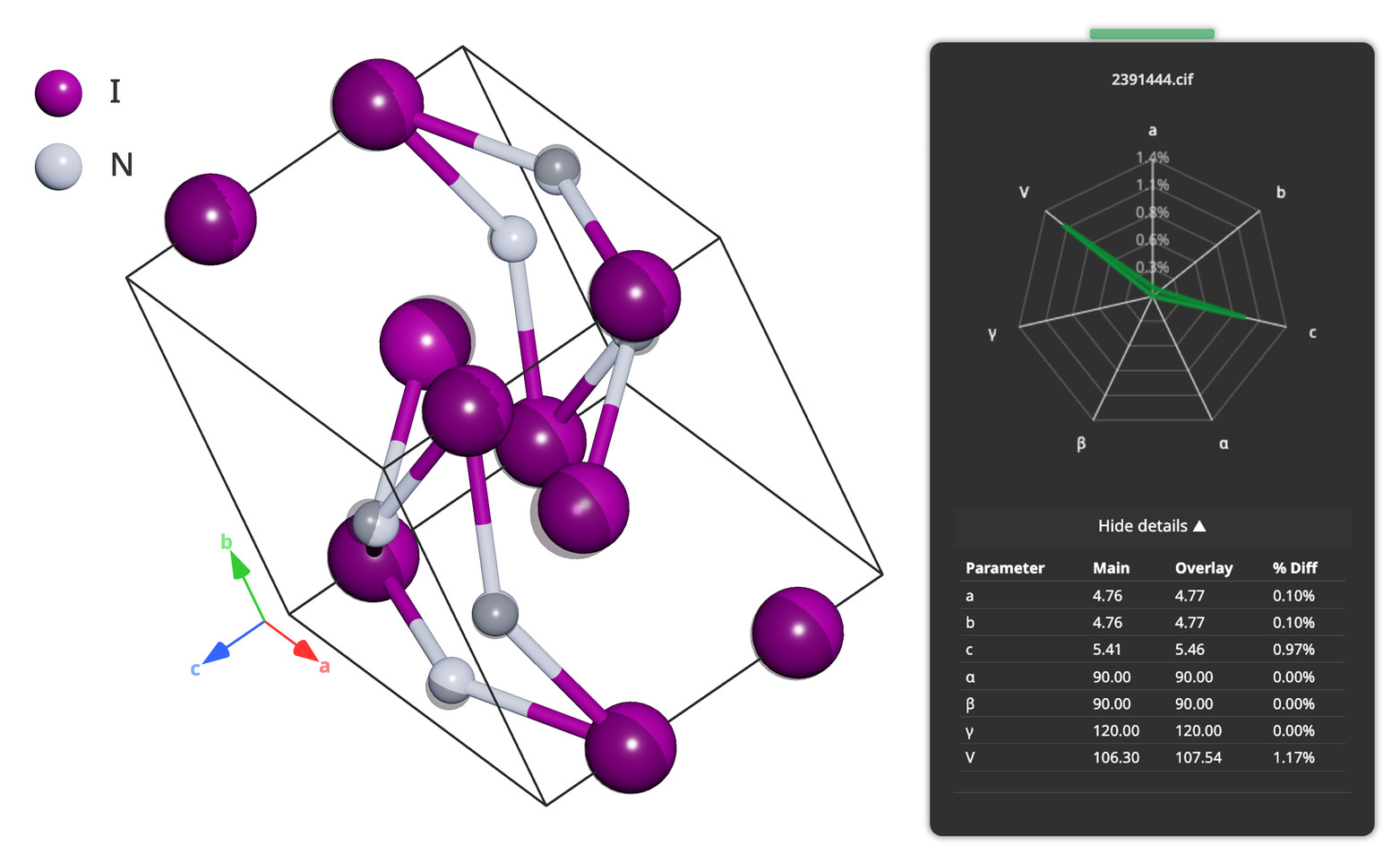}
    \caption{\textbf{Structure Comparison:} The experimentally obtained structure of I$_2$N$_3$ at 95 GPa compared with a variable-cell structural relaxation with Quantum~ESPRESSO from \citet{spender2026high}, with the differences in the cell parameters shown in a radar plot. The darker shaded atoms highlight the positions of the computational result. The experimental structure can be found under CCDC Number: 2391444~\citep{Spender2026_CCDC2391444}.}
    \label{fig:overlay}
\end{figure}

Selecting a second structure in the file browser allows a structure overlay with two-way cross-fading, together with a radar chart of the lattice-parameter differences, for example between a relaxed structure and its (experimental) starting configuration (Fig.~\ref{fig:overlay}). Several structures can be overlaid with per-structure, user-adjustable opacity and an average structure can be created in the file browser. For a side-by-side comparison, several \crysviz tabs or windows in the same browser can be opened, e.g.\ one per structure, and their cameras synchronised: rotating or zooming the structure in one tab rotates and zooms it in all others. Either all tabs show the same view, or each keeps its own orientation and is turned by the same amount. The tabs communicate through a standard browser interface for messages between tabs of the same website.

\subsection{Lattice Analysis} Lattice parameters and angles can be plotted along a trajectory or across a set of structures, e.g.\ a in pressure series against energy, pressure, volume, or a user-defined quantity; selecting a point shows the corresponding structure, and hovering over it reveals more details (Fig.~\ref{fig:lattice}). A similar functionality exists for equation of state analysis.

\subsection{Image and Video Export}

\begin{figure}
    \centering
    \includegraphics[width=1\linewidth]{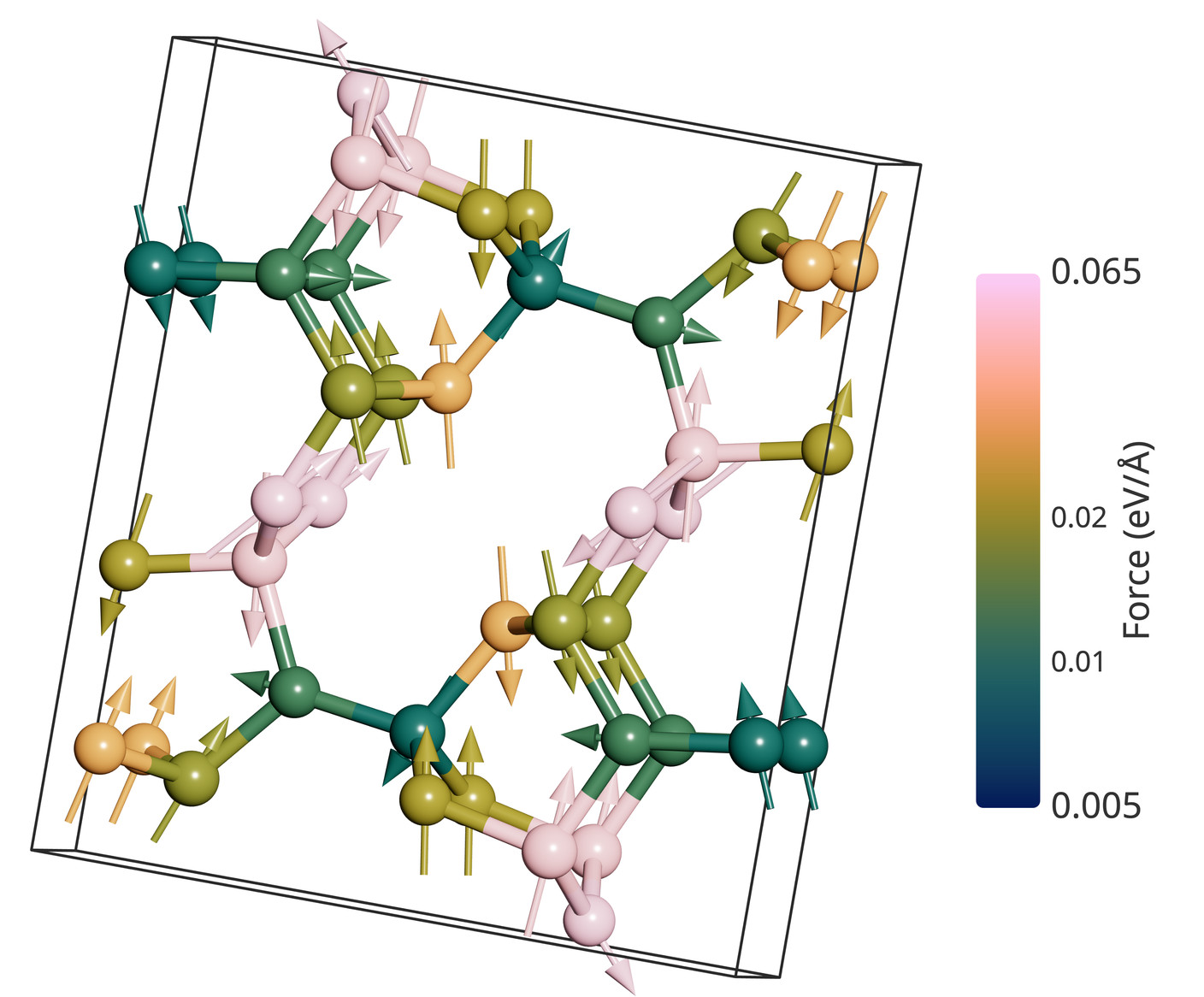}
    \includegraphics[width=1\linewidth]{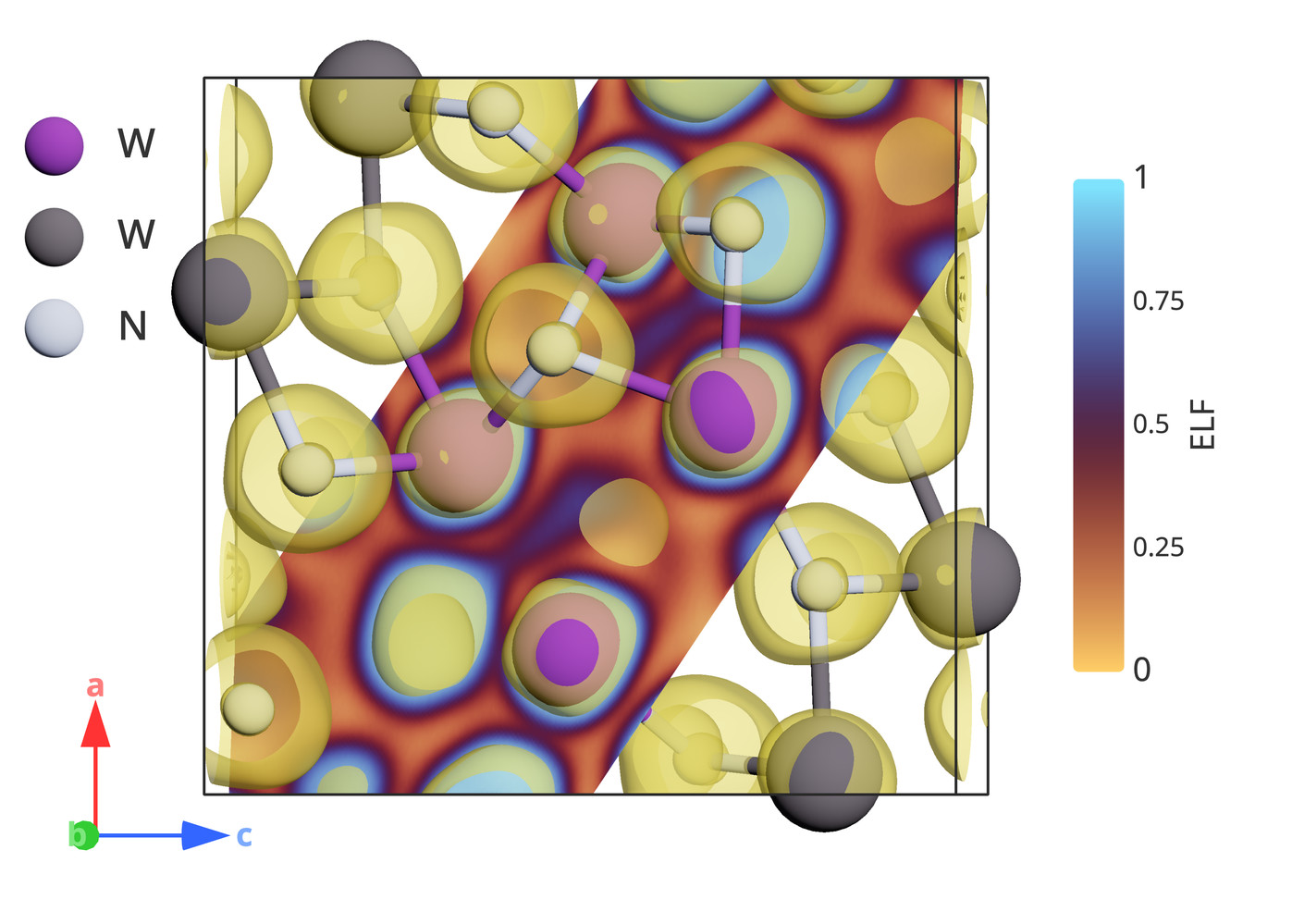}

    \caption{\textbf{Image Export:} \textbf{(top)} Exported snapshot of the trajectory in Fig.~\ref{fig:trajectory}, with forces and colour bar.
    \textbf{(bottom)} Exported image of W$_3$N$_3$ with iso-surfaces of the electron localisation function (ELF) in yellow as well as a plane cutting the four W atoms in purple.
    }
    \label{fig:export}
\end{figure}

The currently visible structure in the structure view, with all colour, bond, and polyhedra settings, force/spin visualisation and colour maps, can be exported as a high-resolution PNG or SVG image. The user can choose the aspect ratio or select the region to export, and whether the background is exported. A chosen region is remembered between exports, so that several steps of a trajectory can be exported at the same orientation and zoom level (Fig.~\ref{fig:export}). Overlays can be exported as well (\textit{cf.} Fig.~\ref{fig:overlay}), and a legend for the atom colours can be included if element-based colouring is used. All rendering modes can be exported as displayed (\textit{cf.} Figs.~\ref{fig:rendering} and \ref{fig:ray-tracing}). All figures that do not show control panels in this manuscript are direct exports from \crysviz.

SVG files can be exported editable: the background, the measurement annotations, the axes, the colour bars and the composition legend are each written as a separate, named group of vector objects (layers in Inkscape). Text is stored as text elements, not converted to outlines or pixels. By default, the structure itself is embedded as a raster image at the chosen resolution, which reproduces every rendering effect exactly as shown on screen. Alternatively, it is exported as vector shapes, with atoms, bonds, unit-cell edges and polyhedron faces as separate, named objects (volumetric fields are so far not supported). A figure can therefore be modified in any vector graphics editor, for example Inkscape or Adobe Illustrator, without rendering it again: fonts, colours and label texts can be changed and legends moved.

\crysviz allows to export videos for presentations directly from the browser.\citep{mediabunny} In \texttt{Showroom Mode}, the structure turns 360$^\circ$ around a chosen axis at an adjustable speed (including a preview mode), and is saved as a seamlessly looping video. There is furthermore a \texttt{Free Recording Mode}, similar to a screen recording, capturing the view while the user rotates, zooms or plays a trajectory/structure. Note, that this mode can record at very high resolution, videos can therefore quickly become large. Videos include the same axes, colour bars and legends as the image exports and are created entirely on the user's computer.  Video export can be combined with ray and path-tracing in \texttt{Showroom Mode}. However, this is very computationally demanding and can depending on the resolution, chosen frame rate and performance of the user's device take several hours.

\subsection{Sharing}
\label{sec:sharing}
\begin{figure}
    \centering
    \includegraphics[width=1\linewidth]{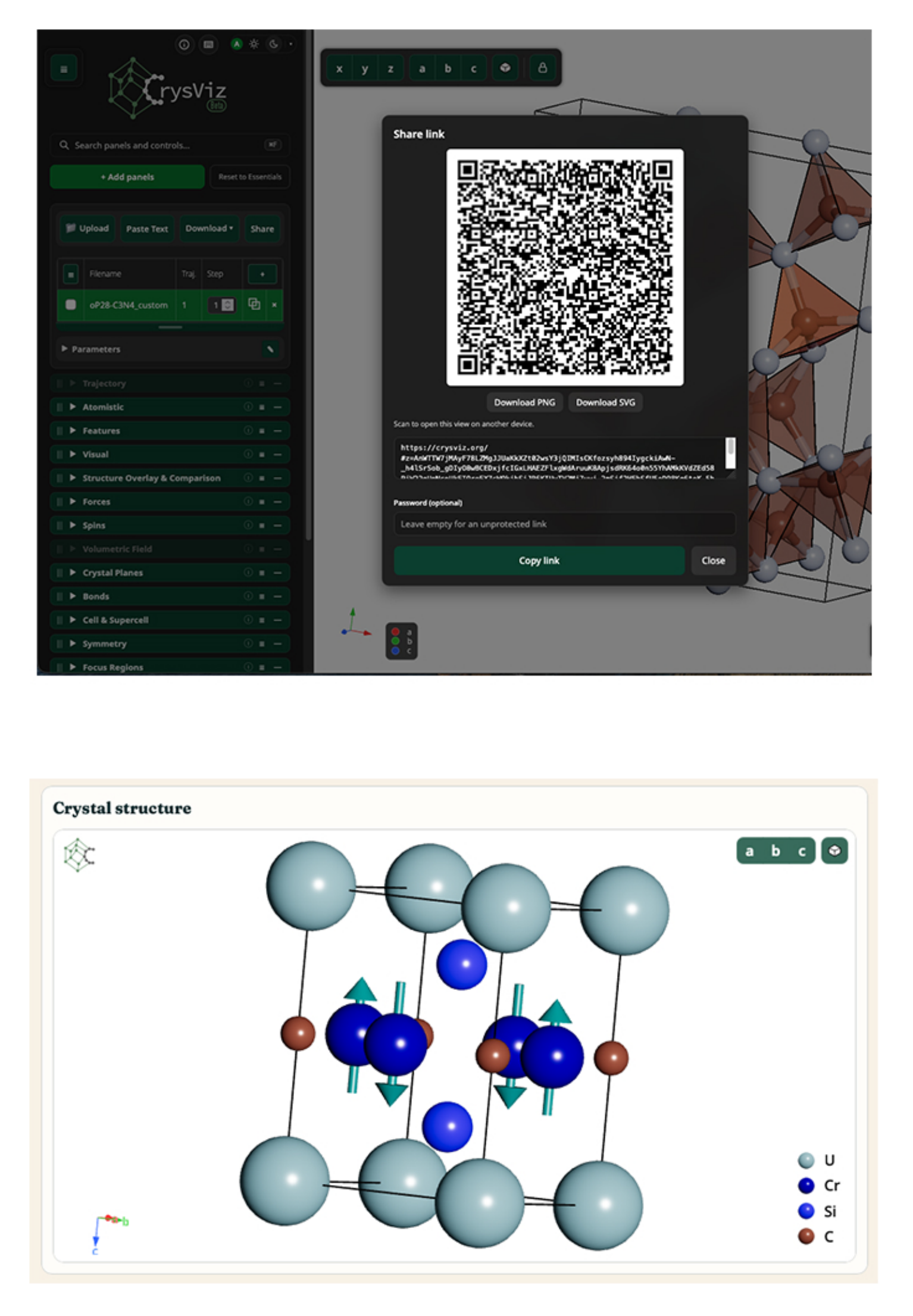}
    \caption{ \textbf{Sharing and Embedding:} \textbf{(top)} Share dialogue with a link, QR code, and the password field that allows optional encryption of the share link. \textbf{(bottom)} \crysviz embedded as a widget in the Anyterial Altermagnets database \citep{sufyan2025_high_throughput} (example UCr$_2$Si$_2$C, \texttt{anyt.am-1-5}~\citep{anyterialExample}).}
    \label{fig:sharing}
\end{figure}

The complete session, including all trajectory frames, volumetric data, styling, camera, and measurements, can be saved in a single \texttt{.crysviz} file. The current view, comprising the structure, its styling (bonds, polyhedra, colours, etc.), the camera orientation, and any measurements, can also be captured in a shareable link or QR code. The link itself contains the information to regenerate the view. The information travels in the link, which browsers do not send to a server. Only for an unchanged structure from a supported database (e.g.\ Alexandria), the link stores just a reference, and the structure is fetched from that database. Links can be encrypted with a password, so that only chosen recipients may open them (Fig.~\ref{fig:sharing}).

\subsection{Agentic Interoperability}

\crysviz can also be driven by AI agents through its Python package. With \texttt{pip install "crysviz[mcp]"}, the package provides the option that users can run a local server for the Model Context Protocol (MCP)~\citep{anthropic2024mcp}, the open standard through which AI assistants call external tools. Any assistant that supports MCP (e.g.\ Claude, Codex, GitHub Copilot, or a local model with an MCP-capable client) can control the same \crysviz window as the user. The agent can load and edit structures, find atoms, bonds and polyhedra, read the symmetry, run relaxations and MD, and retrieve images of the current view, while the user interacts with the same structure. These tools are general building blocks that the agent combines for a task. For example, asked to ``put a sulfur atom at the centre of every six-membered carbon ring'', the agent exports the structure with the \crysviz bond rule, finds the rings in its own script (or with locally installed Python packages such as ASE or pymatgen), adds the S atoms at the ring centres, and checks that the space group is kept. The \crysviz MCP server runs on the user's device, so no structure leaves it except through the LLM agent the user has chosen. Using a local model, nothing leaves it at all.

\section{On-device machine-learned interatomic potentials}
\label{sec:mlip}

\crysviz not only \emph{displays} trajectories but can also \emph{compute} them on the device, using machine-learned interatomic potentials compiled to WebAssembly. Relaxations and MD runs are launched from the viewer, and the frames are streamed live into the trajectory player. A completed run is a fully browsable trajectory (Section~\ref{sec:trajectories}). An exported \texttt{.crysviz} file contains the full trajectory (per frame: lattice, positions, forces, energy and stress) and the name of the potential that computed it; for an MD run also the temperature, target temperature, total energy and pressure of every frame, and the velocities of the last frame. It keeps the visual settings, and the run settings (time step, thermostat, ensemble, number of steps, calculator).

\subsection{Potentials}

\crysviz provides two tiers of potentials that run on the device, and can additionally control calculations running in a user's Python backend. The primary built-in potential is a neuroevolution potential (NEP), whose evaluation code (\textsc{NEP\_CPU}, part of \textsc{GPUMD}) runs in the browser~\citep{Fan2017,Fan2022GPUMD} as a WASM module. NEP attains good accuracy at very low computational cost per evaluation~\citep{Fan2021NEP,Fan2022MultiNEP}. The pre-trained NEP89~\cite{liang2026nep89} weights are loaded on-the-fly, allowing structural relaxations and MD runs for a wide range of chemistries.

Secondly, PET-MAD~\citep{Mazitov2025PETMAD} potentials can be used directly in \crysviz. A WASM module derived from \textsc{mlip.cpp}~\citep{mlipcpp} loads potentials in the GGUF format (provided on Hugging Face). Other models in this format, including user fine-tuned ones, can be loaded from a link to a repository. By default, two PET-MAD models are available (XS and S), with multi-threading and WebGPU support. The XS model is about 10 and the S model about 30 times more computationally demanding than NEP89, but provides a higher accuracy, especially at elevated pressures.
The S model is already on the verge of what can be run efficiently in a browser. For larger models, the Python backend (see below) should be run by the user.

While MLIPs are approaching a state where they can be considered a true replacement for DFT calculations over a wide range of chemistries, pre-trained models still come with limitations. No guarantee can be given that the models available in \crysviz produce correct results, especially at high pressure or for complex chemistries. As with any use of MLIPs, careful testing and, in most cases, fine-tuning to the particular system are highly recommended. As a rule of thumb, the low computational cost NEP model provides good results at ambient conditions for equilibrium structures and can in particular be used for quick testing, but has clear limitations in finding equilibrium structures at pressures above 5--10~GPa. PET-MAD XS is therefore recommended for all calculations involving pressures above 5~GPa. \crysviz displays warnings when MLIPs are used to remind the user of these limitations. Starting simulations for too large cells is blocked, as memory safeguards are strongly browser dependent: with the current implementation, 15k atoms can theoretically be run for NEP and 1000 for PET-MAD. These limits are generous: simulations of this size require a multi-core system or a GPU, to reach reasonable steps/s.

To go beyond these limits, or when a particular electronic-structure code or higher-tier MLIP is required, a calculation can be delegated to an external calculator compatible with the Atomic Simulation Environment (\textsc{ASE})~\citep{Larsen2017}, running in the user's Python environment and controlled via the \crysviz interface. Example scripts are provided that allow users to set up their own local server. The optimiser or MD integrator runs in \crysviz and requests energies, forces and stresses from the backend, so that each step appears in the trajectory player. This allows, for example, the usage of MACE foundation models~\citep{Batatia2022MACE,Batatia2023Foundation}, or any ASE-compatible code (VASP, Quantum~ESPRESSO, CASTEP, FHI-aims or another machine-learned potential). Trajectories are stored in \crysviz (the provided server backend can be extended for immediate local storage) and can make direct use of all visualisation and analysis capabilities. Example sscripts for different MLIPs can be found in the GitHub repository. This option is particularly useful for prototyping. For production simulations, it is recommended to run simulations directly, e.g., via ASE, and load the trajectory later into \crysviz. It is possible to run the server on an HPC system and forward the port to the local machine. How well this setup performs depends strongly on the network connection. Users should check in the guidelines for the particular HPC system if this is permitted. 

\subsection{Relaxation, Molecular Dynamics, Ordered Structure Generation and Single-Point Evaluation}
\label{sec:mlip-runs}

The following functionalities use either the NEP89 or the PET-MAD potentials.

Single-point evaluations give the energy (per atom), forces, and the stress tensor (pressure) of the currently displayed crystal structure. Geometry optimisations can be performed including variable-cell relaxations to a target pressure, with the standard controls (step limit, force and stress tolerances, and target pressure)~\citep{Bitzek2006}. Molecular dynamics simulations can be performed in the NVT ensemble with a stochastic velocity-rescaling thermostat~\citep{Bussi2007}, or in the NPT ensemble with an additional stochastic cell-rescaling barostat~\citep{Bernetti2020} (so far only isotropic rescaling is implemented). An optional simulated-annealing scheduler can be used to test structural stability. A relaxation always starts a new trajectory, leaving the original unchanged, while an MD run can be continued from the end of an earlier run. During a simulation, the trajectory plot (Section~\ref{sec:trajectories}; Fig.~\ref{fig:trajectory}) updates in real time; the maximum force and, for MD, the potential and kinetic energies are reported as text. If the Wyckoff mode is active, relaxations and MD runs preserve the space group (transitions to supergroups are always allowed).

For a crystal structure with fractional occupations, the \emph{Order Structure} tool can construct ordered approximations (a stoichiometry-preserving first-neighbour-shell optimisation, or a majority-occupancy rule), and different ordered versions can be automatically ranked by their energy. This simple model is not yet a full implementation of special quasirandom structures (SQS)~\citep{Zunger1990}; for rigorous alloy statistics, dedicated tools should be used. The generated cells can, however, help to investigate ways to resolve fractional occupations or give first approximations for alloys that can later be refined. Finally, equation of state calculations (Section~\ref{sec:modules}) can be performed using the MLIPs.

\section{Customisation}
\label{sec:custom}

\crysviz is extensively customisable, both per structure and through persistent defaults. Colours can be assigned per species (using built-in or user-defined colour tables), per atom, or according to a numerical quantity (force magnitude or bond length), using a range of colour maps including perceptually uniform, colour-vision-safe Scientific colour maps~\citep{Crameri2018,Crameri2020} in addition to standard Viridis and Plasma maps. Colour bars are generated from eight predefined colour maps, with user-editable labels and automatically detected or user-defined ranges (linear or logarithmic). They can be placed anywhere in the structure view, flipped between vertical and horizontal, and resized, and their labels can be flipped between left/right and top/bottom. Colour bars are included in exported images. Bonds are defined by element-pair cutoffs, either via the default values or a user-supplied table. Atomic radii, default colours, cutoffs, and fully rebindable keyboard shortcuts can be overridden globally through an editor and saved to a settings file (Fig. \ref{fig:user-settings}). Several colour themes with light and dark variants are available for the interface.
\section{Specialised analysis modules}
\label{sec:modules}

\crysviz is structured so that additional, self-contained analysis modules can be built on top of the core
viewer. Each module reads the current structure, computes its own results, and displays them in a linked
window that highlights the relevant atoms in the 3D view. A small add-on interface is available, and the manual has a guide to writing new modules. Currently, three existing modules extend \crysviz for more project or domain specify applications:

\paragraph{Equation of State.} A second- or third-order Birch--Murnaghan~\citep{Birch1947} or Vinet~\citep{vinet1987compressibility} equation of state
can be fitted to energy--volume or pressure--volume data (Fig.~\ref{fig:EOS}). The data may be loaded from a single test file (P-V, E-V or E-V-P columns; no structures are shown), a single or set of trajectories, e.g., variable-cell structural relaxations, a \texttt{.crysviz} file or generated entirely on the device: the module relaxes the current cell at a series of target pressures using a chosen potential (Section~\ref{sec:mlip}) and collects the resulting volumes and energies. The fit returns the equilibrium volume, the bulk modulus, and its pressure derivative. The module furthermore plots the fitted $E(V)$, and $V(P)$ or $P(V)$ curves against the input data. Reference data, e.g.\ from experiment, can be overlaid for comparison, and deviations in volume at a given pressure are reported in \%. When EOSs are combined or loaded to a \texttt{.crysviz} trajectory, several EOSs can be shown in the same plot.  \crysviz warns the user if the EOS has jumps. Details on the EOS analysis features can be found in the manual.

\paragraph{Energy Landscapes.} A second module visualises symmetry-restricted energy landscapes~\cite{parackal2026symmetry} (Fig. \ref{fig:energy_landscapes}): the
energy of a structure computed over a grid of symmetry-adapted atomic displacements, evaluated by several energy models, both machine-learned and DFT, in parallel. Read from a precomputed file, each model is shown as a colour map relative to its own minimum; selecting a point within a map displaces the atoms in the 3D view in real time, and two compatible landscapes can be linked along a shared coordinate. The module thus allows an interactive, on-device comparison of how different models describe the same distortion. It also illustrates how a module can introduce its own analysis window, file format, and coupling to the 3D view while reusing the remainder of \crysviz.

\paragraph{Phonons.} A third module allows the visualisation of phonon calculations from \textsc{phonopy}~\citep{Togo2023}. Band structures and densities of states are plotted in a linked window (Fig. \ref{fig:phonons}. Selecting a point in the band structure starts an animation of the corresponding mode on the atoms in the 3D view, optionally with displacement arrows. For points outside of the $\Gamma$ point, a commensurate supercell is automatically suggested. Imaginary modes are listed separately.

\section{Availability}
\label{sec:availability}

\crysviz is free and open-source software, released under the GNU Affero General Public License v3.0 or later (AGPL-3.0-or-later).  \crysviz is available under \url{crysviz.org}. The source code, issue tracking, and build/deployment instructions are available at \url{https://github.com/CrysViz/crysviz}. Third-party components used by \crysviz, including \textsc{Moyo}~\citep{Shinohara2025}, the NEP\_CPU inference engine and NEP potential weights~\citep{Fan2022GPUMD}, PET-MAD~\citep{Mazitov2025PETMAD}, \textsc{mlip.cpp}~\citep{mlipcpp}, the \textsc{httk} space-group data~\citep{httkdata}, and \textsc{three.js}~\citep{ThreeJS}, are attributed and separately licensed as documented in the repository.

\section{Conclusion}

\crysviz natively reads common crystallographic formats, such as CIF and magnetic CIF, as well as the input and output files of simulation codes such as VASP, Quantum~ESPRESSO, CASTEP, and FHI-aims. Volumetric data, e.g.\ charge densities, and phonon calculations from \textsc{phonopy} are supported as well. Relaxations and MD runs are shown as trajectories, together with their forces and magnetic moments. New trajectories can be computed within the viewer using machine-learned potentials on the user's device. In the Wyckoff mode, these runs keep the symmetry of the structure.

\crysviz also offers new ways of sharing data. A complete view, including styling and measurements, fits into a compact link or QR code that can be encrypted with a password. Unpublished structures can thus be shown to collaborators, and a figure in a paper or on a poster can link to its interactive view. The same viewer can also be embedded in other web pages or controlled from Python.

Finally, users can add their own analysis modules through a small add-on interface, as the energy-landscape module illustrates. \crysviz is not meant to replace established desktop tools. Instead, it complements them and can grow with the needs of its users.

\section*{Acknowledgements}

We thank Amanda Ehn , Timothy H. Lees (Theoretical Physics Division at LiU) and Dominique Laniel (University of Edinburgh) for discussions and input during the development of \crysviz. We would like to furthermore thank Adrian Kjellén, Oscar Groppfeldt, Ashis Kundu, Joakim Stenborg, Davide G. Sangiovanni, (Theoretical Physics Division at LiU), Yuqing Yin, Gerd Steinle-Neumann (Bayerisches Geoinstitut, University of Bayreuth), Hannah Shuttleworth, Xuejing He, Sarah Bolton, and Qian Zhang (University of Edinburgh), and Kieran Bozier (University of Cambridge) for their help during the beta test.

F.T., A.S.P, and H.t.E acknowledge support through ERC Grant (UNMASCC-HP,  10.3030/101117758). Funded by the European Union. Views and opinions expressed are however those of the author(s) only and do not necessarily reflect those of the European Union or the European Research Council Executive Agency. Neither the European Union nor the granting authority can be held responsible for them. F.T. acknowledges furthermore support from the Swedish government's Strategic Research Area in Materials Science on Functional Materials at Linköping University (Faculty Grant SFO-Mat-LiU 2009-00971). F.T. and A.S.P acknowledge support by the Wallenberg Initiative Materials Science for Sustainability (WISE) funded by the Knut and Alice Wallenberg Foundation.

O.B.L. acknowledges the support of the European Union under Horizon Europe for the QRC-4-ESP project (Grant Agreement 101129663) and the QUEST project (Grant Agreement 101156088).

 R.A. acknowledges the Swedish e-Science Research Centre (SeRC).

Some of the computational data visualised in this paper is the result of projects enabled by computational resources provided by the National Academic Infrastructure for Supercomputing in Sweden (NAISS), funded by the Swedish Research Council.

\section*{ Use of AI}

\crysviz was developed with the aid of tools based on large language models (LLMs), but the resulting software is the product of nearly a year of iterative, human-directed development. The authors conceived the software architecture, functionality and design; directed and evaluated the development process; selected, modified and integrated generated code; and extensively tested, debugged and refined the resulting implementation. The authors have also made extensive internal use of the software throughout its development. The text of this paper was written by the authors with the assistance of LLM-based tools for drafting, review and editing.

\bibliographystyle{apsrev4-2}
\bibliography{references}

\clearpage
\onecolumngrid
\setcounter{figure}{0}
\renewcommand{\thefigure}{S\arabic{figure}}
\renewcommand{\theHfigure}{S\arabic{figure}}
\setcounter{page}{1}
\renewcommand{\thepage}{S\arabic{page}}
\begin{center}
    {\Large\bfseries Supplementary Material}\\[6pt]
    {\large \crysviz: A Portable, Browser-Based Tool for Crystal Structure Visualisation and Analysis}
\end{center}
\vspace{6pt}

The supplementary material consists of a range of additional screenshots and exported images highlighting further \crysviz functionalities.
\suppressfloats[t]

\begin{figure*}
    \centering
    \includegraphics[height=9cm]{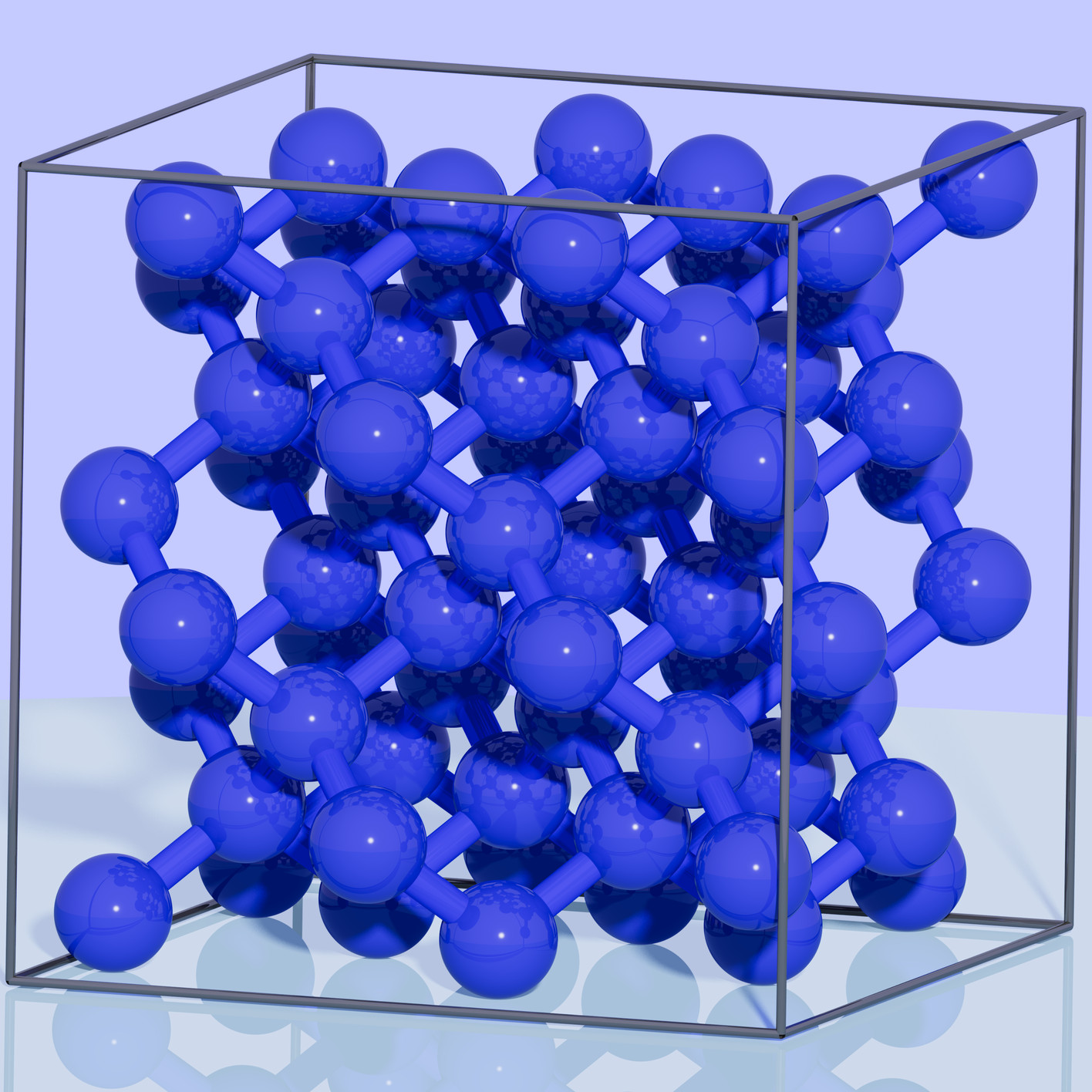}
    \includegraphics[height=9cm]{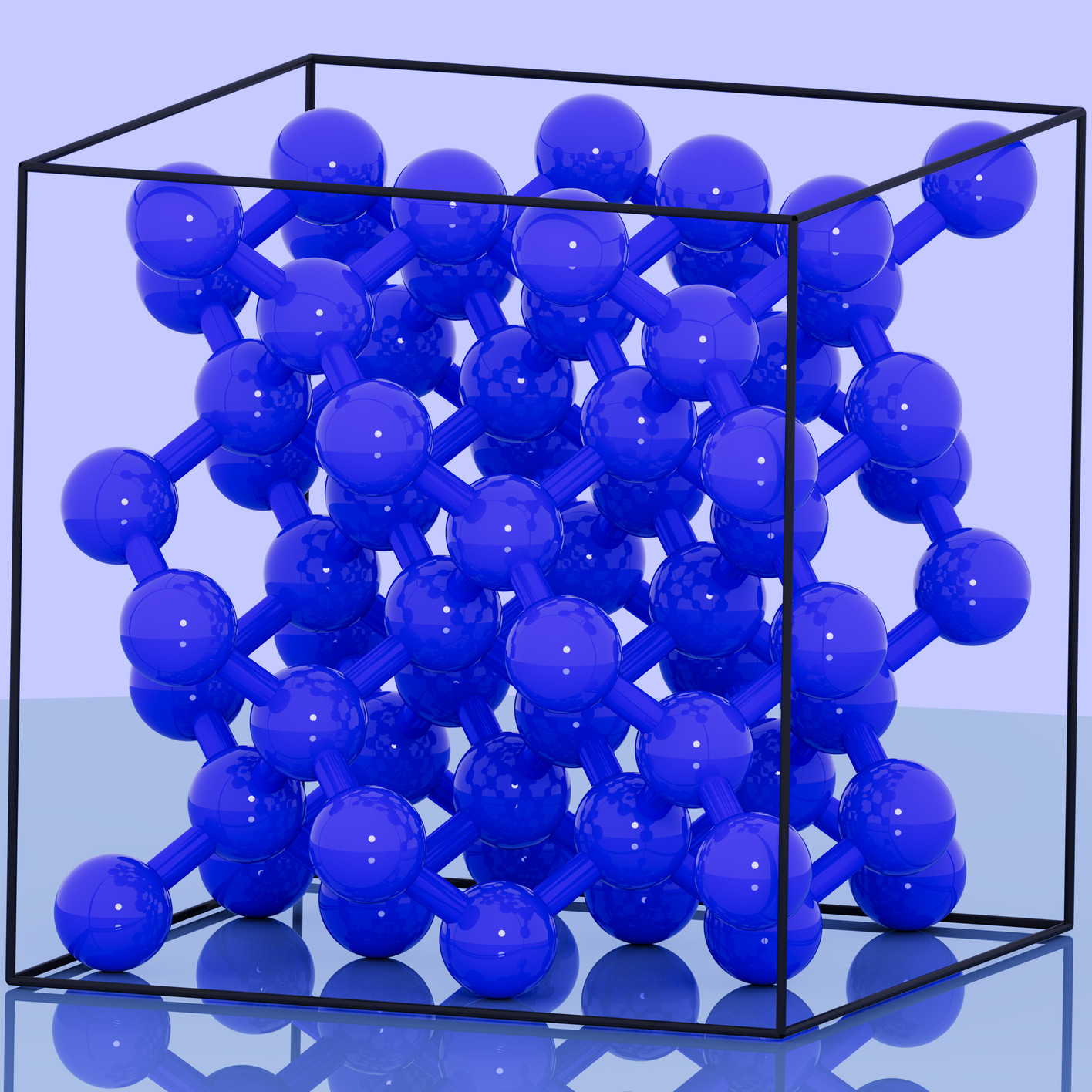}
    \caption{\textbf{Ray and path tracing:} The two images compare ray tracing (top) and path tracing (bottom) for silicon (Si). Path-tracing is a more expensive rendering technique, but allows more realistic representation of some materials such as the glossy atoms shown here. On a Linux system with an NVIDIA RTX 5060 Ti a full quality $3840\times3840$ render of the top figure takes around 15 seconds, and of the bottom one 16 minutes (although the image renders gradually and is at nearly final quality already at around 1 minute).}
    \label{fig:ray-tracing_comp}
\end{figure*}

\begin{figure}
    \centering
    \includegraphics[width=0.8\linewidth]{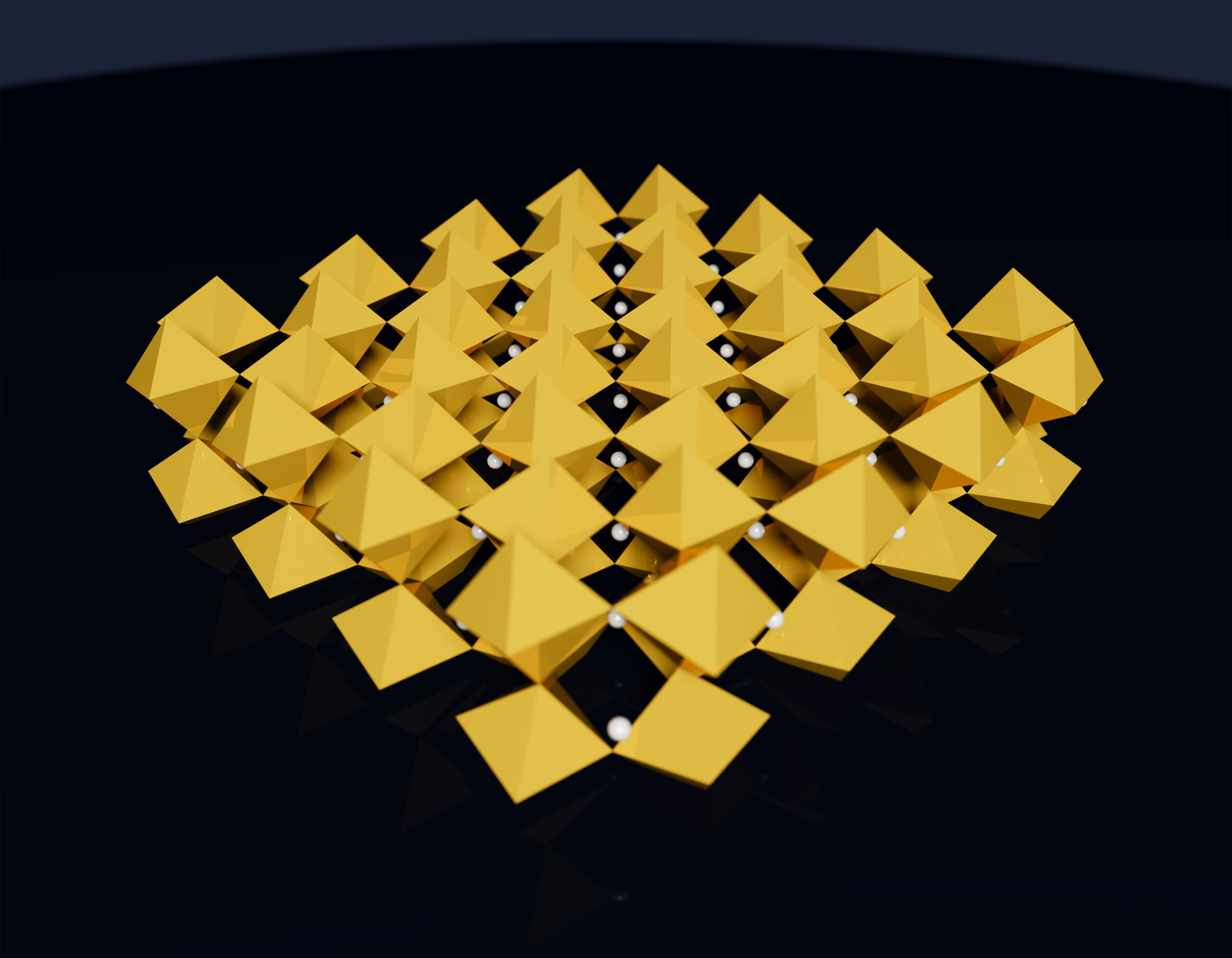}
    \hfill
    \includegraphics[width=0.8\linewidth]{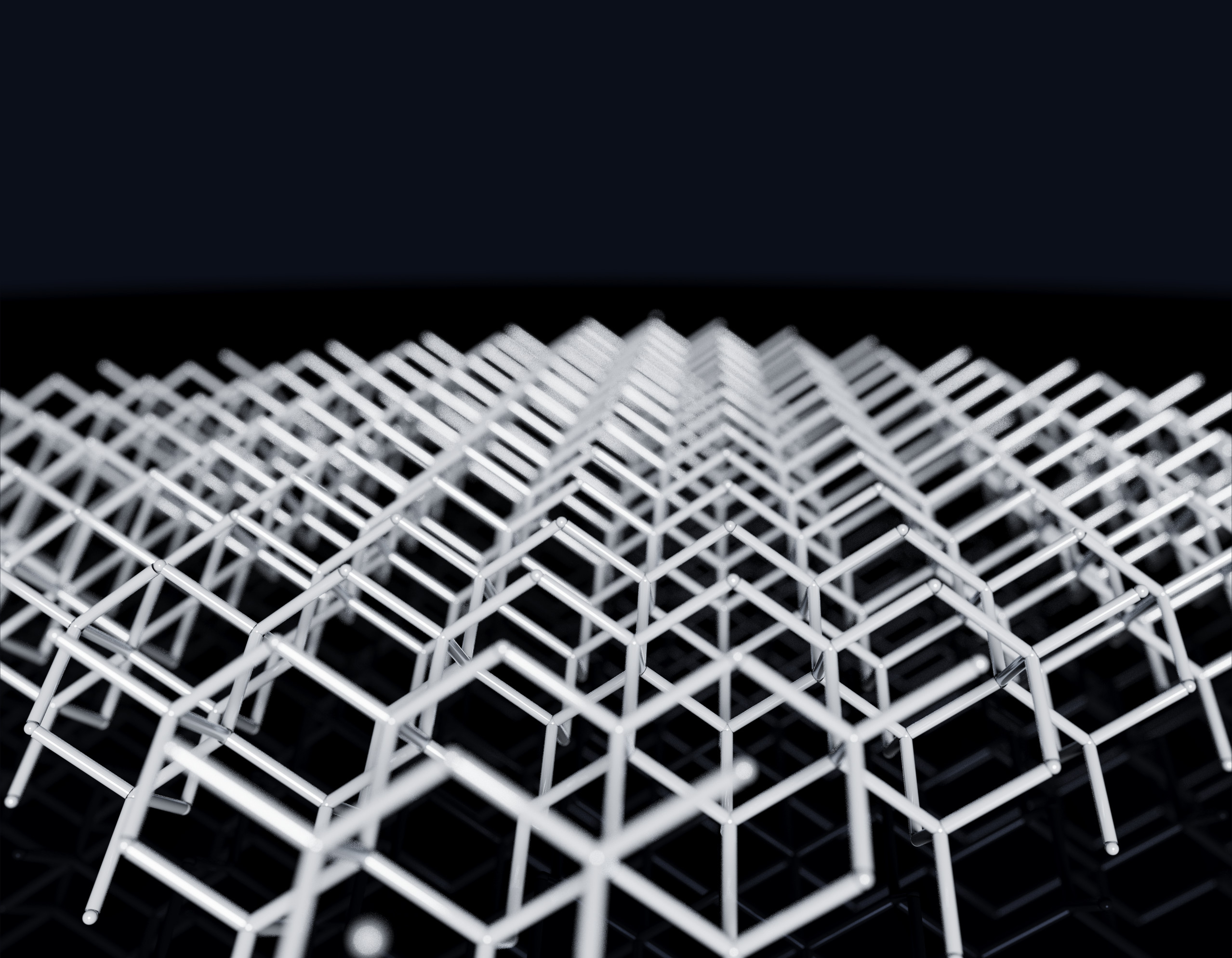}
    \caption{\textbf{Cover Images:} The advanced rendering styles (ray- and path-tracing) allow to export high resolution images beyond simple structure visualisation that can be used in presentations, for websites, or as cover images.}
    \label{fig:cover_style}
\end{figure}

\begin{figure}
    \centering
    \includegraphics[width=\linewidth]{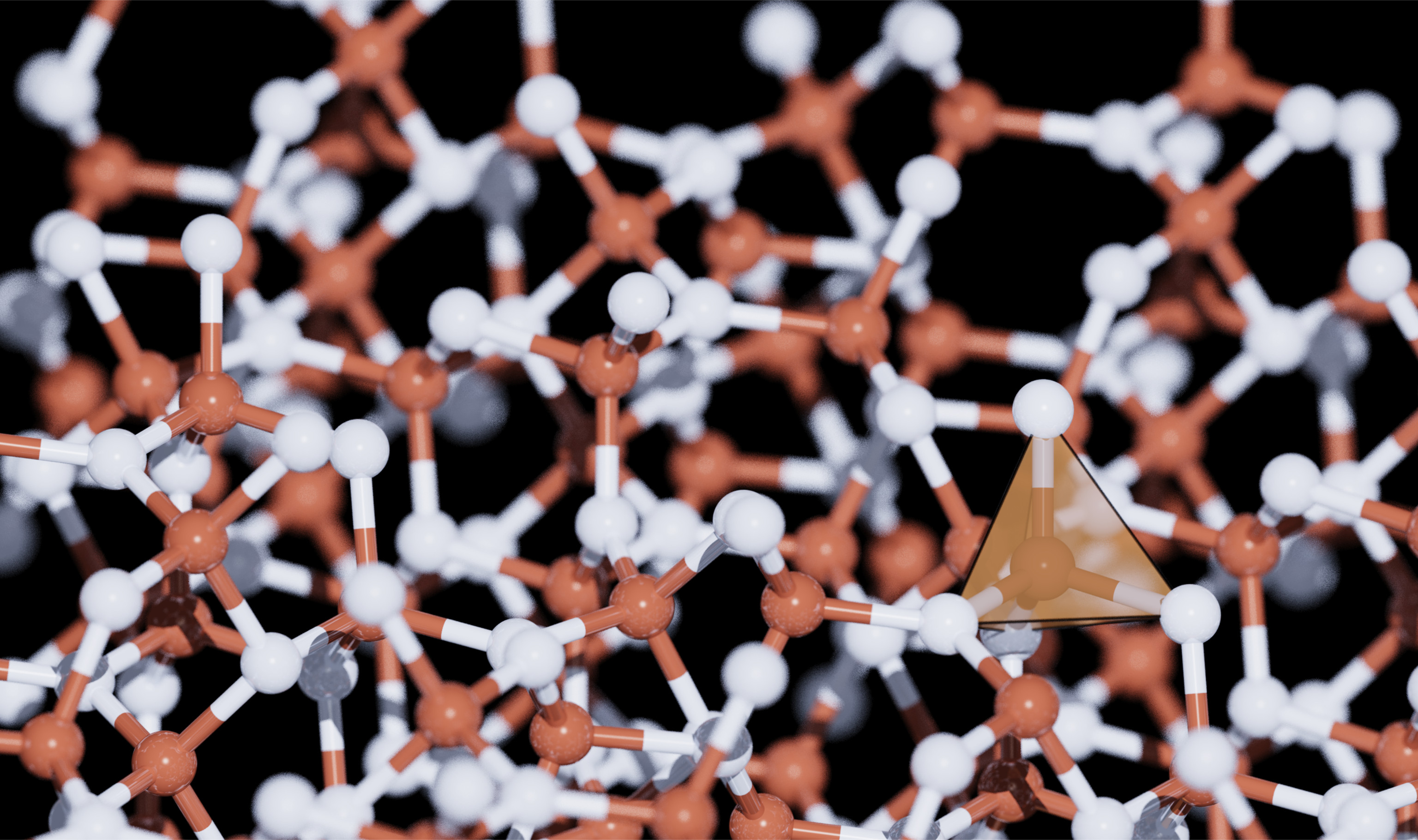}
    \includegraphics[width=\linewidth]{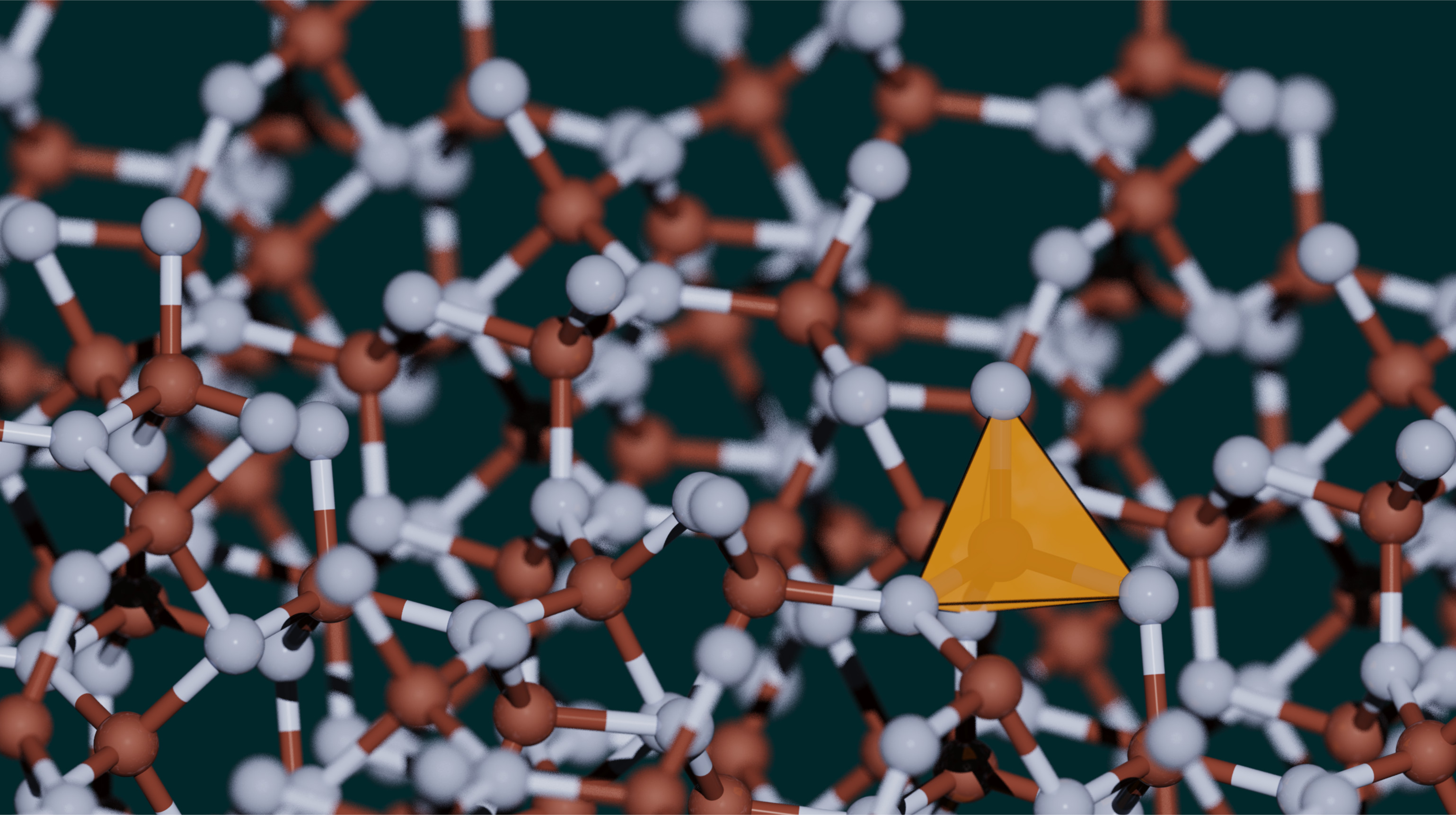}
   \caption{\textbf{Lighting, Shadows, Reflectivity and Backgrounds:} Ray- and path-tracing allow a range of settings for the lighting, shadows, reflectivity and texture of atoms, bonds and polyhedra. The lighting furthermore reacts to the background colour. The lighting is here chosen less soft compared to Figure \ref{fig:ray-tracing} leading to harsher shadows from bonds and atoms.}
    \label{fig:Aperature}
\end{figure}

\begin{figure}
    \centering
    \includegraphics[height=0.5\linewidth]{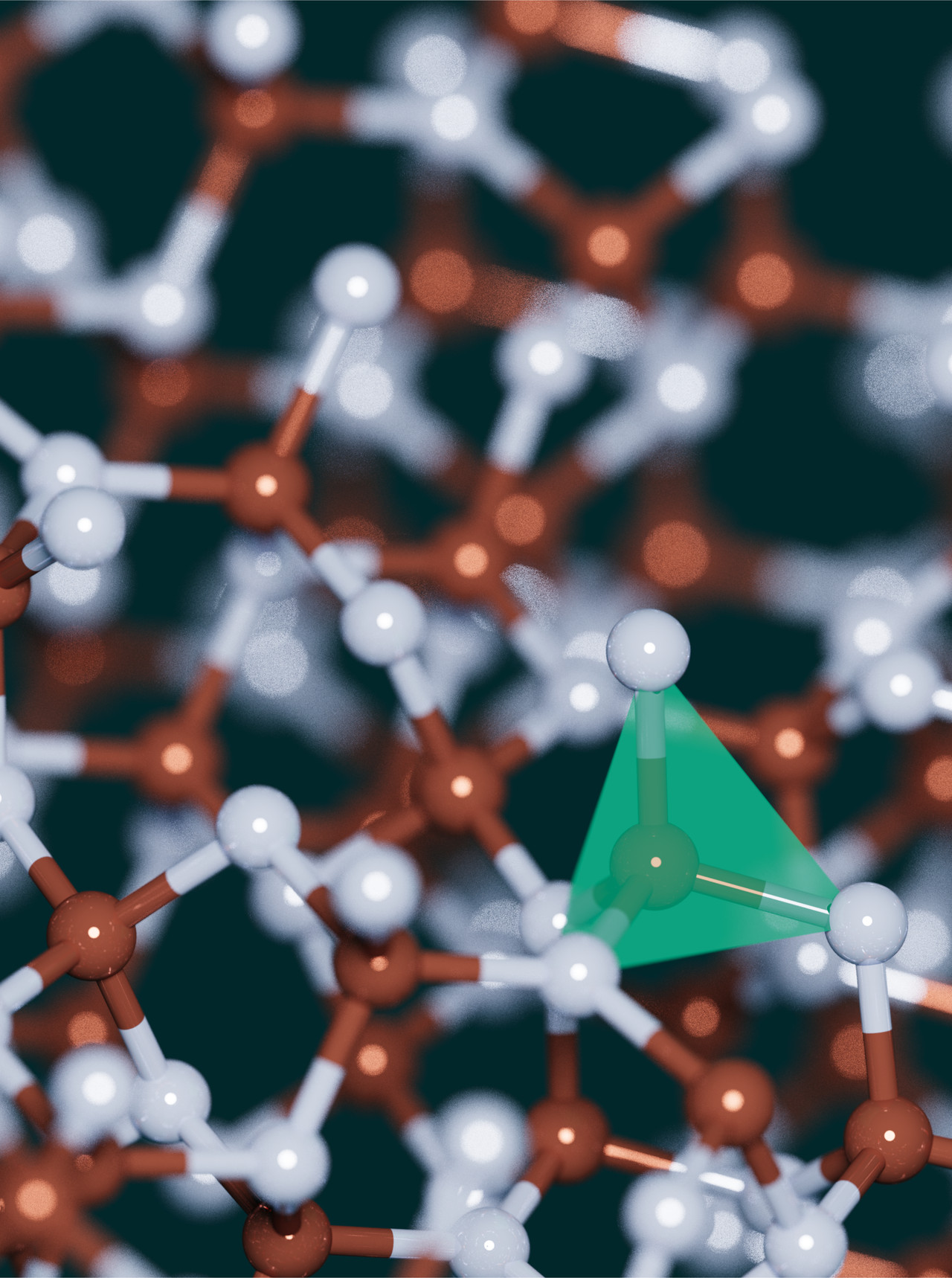}
    \includegraphics[height=0.5\linewidth]{plots/oP28-C3N4_6_13.jpg}
   \caption{\textbf{Aperture:} Ray- and path-tracing feature the option chose an aperture and focus distance that allows to create more artistic visualisations with fore- and/or background out of focus.}
    \label{fig:Aperature2}
\end{figure}

\begin{figure*}
    \centering
    \includegraphics[width=1\linewidth]{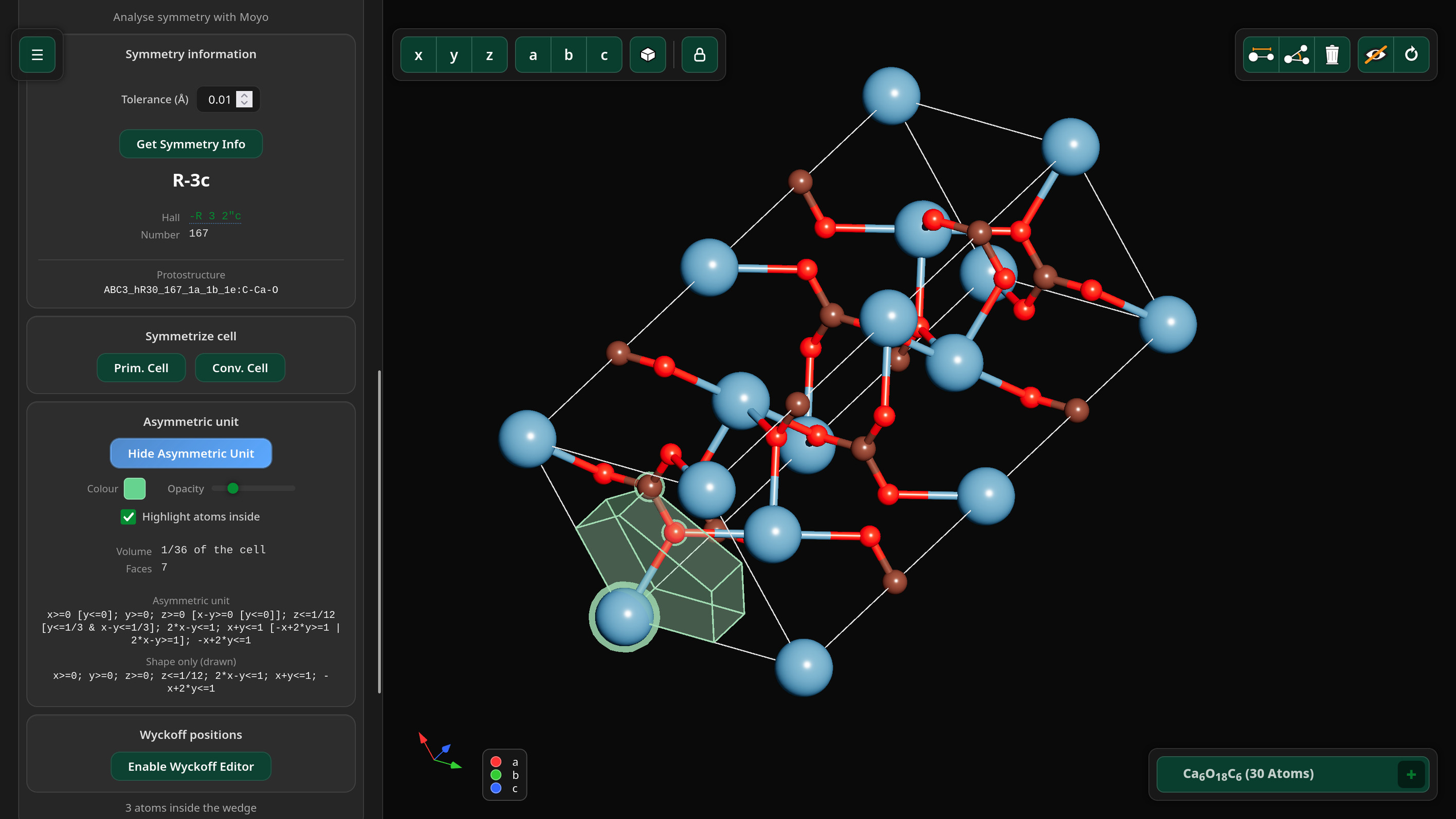}
\includegraphics[width=1\linewidth]{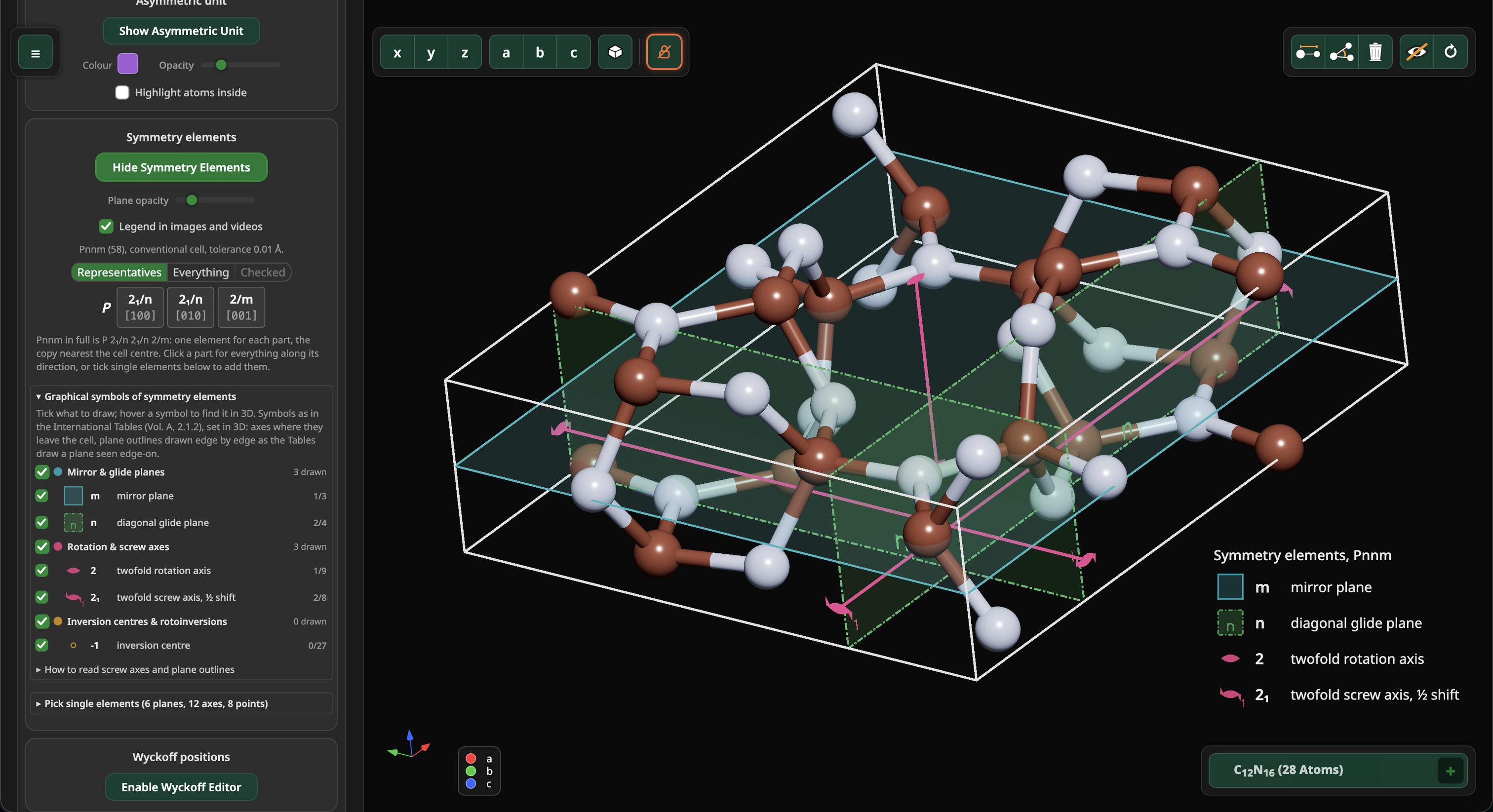}
    \caption{\textbf{Asymmetric Unit and Symmetry:} \textbf{(top)} \crysviz allows visualisation of  real-space asymmetric units (ASU).  The example shows calcite, CaCO$_3$ (\citet{Maslen1993calcite} (COD 2100992; space group $R\bar{3}c$, No.~167)), with the ASU highlighted in green. The structure is shown in the conventional hexagonal cell. The Symmetry panel (left) lists the space group, the volume of the ASU (1/36 of the cell) together with the inequalities that define it. The atoms inside the asymmetric unit are highlighted. Ca--Ca bonds are switched off. \textbf{(bottom)} Symmetry elements of \textit{oP}28-C$_{3}$N$_{4}$ in space group $Pnnm$ (No.~58, full symbol $P\,2_1/n\,2_1/n\,2/m$), shown in the conventional cell for the $[100]$ part of the symbol. Drawn are the four $2_1$ screw axes $\parallel\mathbf{a}$ (pink, with the  International Tables' screw symbol where each leaves the cell), the two $n$-glide planes $\perp\mathbf{b}$ at $y=\tfrac14,\tfrac34$ (green, glide $\tfrac12(\mathbf{a}+\mathbf{c})$) and the mirror planes $\perp\mathbf{c}$ at $z=0,\tfrac12$ (cyan).For clarity, the $n$-glide planes $\perp\mathbf{a}$, the 2-fold axes $\parallel\mathbf{c}$, the $2_1$ axes $\parallel\mathbf{b}$ and the inversion centres are not shown; symmetry was found with a tolerance of 0.01\,\AA.  
    }
    \label{fig:asu}
\end{figure*}

\begin{figure*}
    \centering
    \includegraphics[width=1\linewidth]{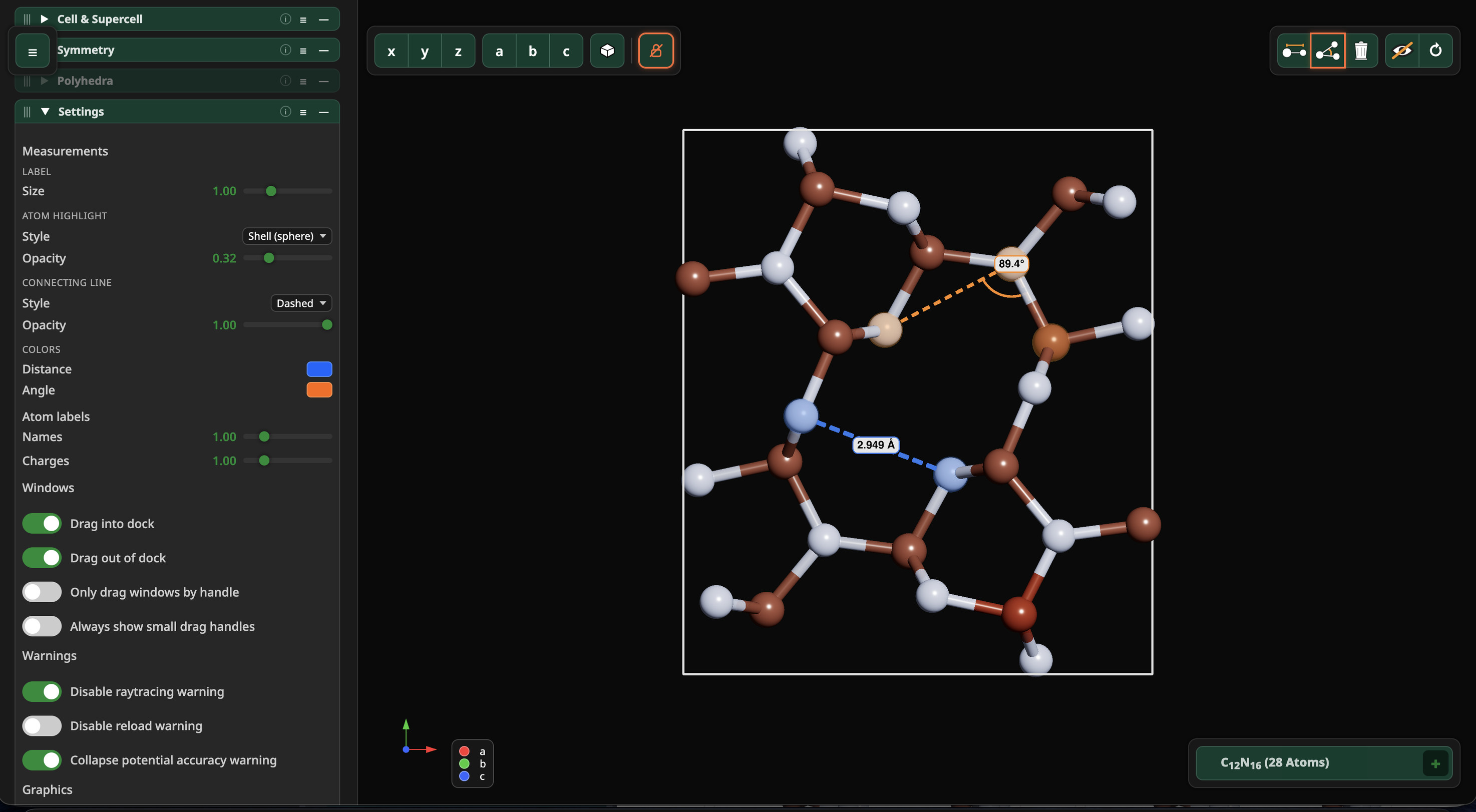}
    \caption{\textbf{Measurements:} \crysviz allows measurement of distances and angles. The label sizes, colours, and highlight style can be chosen by the user. Measurements can be included in the image export and update live when a trajectory is played, during a relaxation, MD or when switching between steps. }
    \label{fig:Measurements}
\end{figure*}

\begin{figure*}[t]
    \centering
    \includegraphics[width=0.48\linewidth]{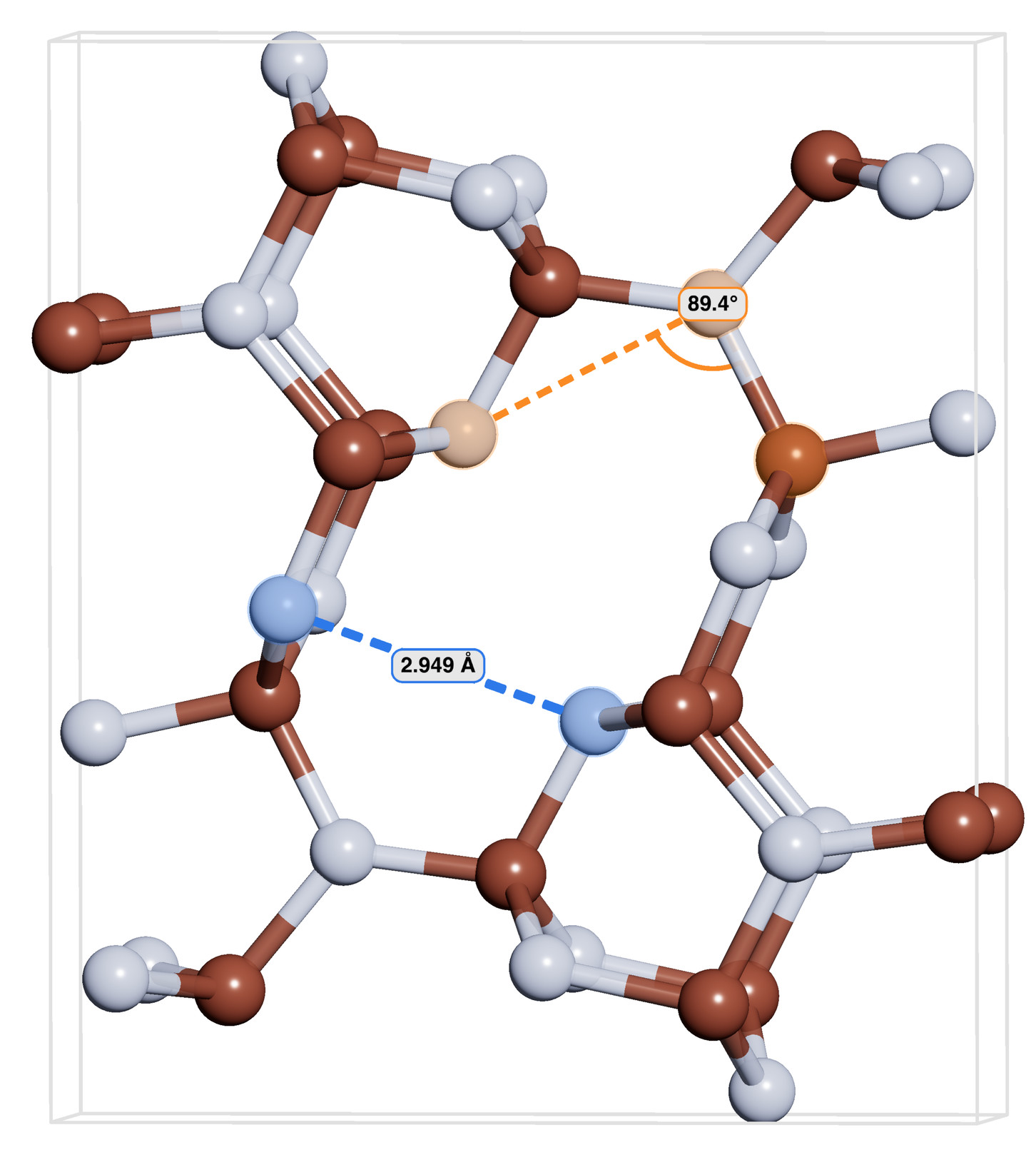}%
    \hfill
    \includegraphics[width=0.48\linewidth]{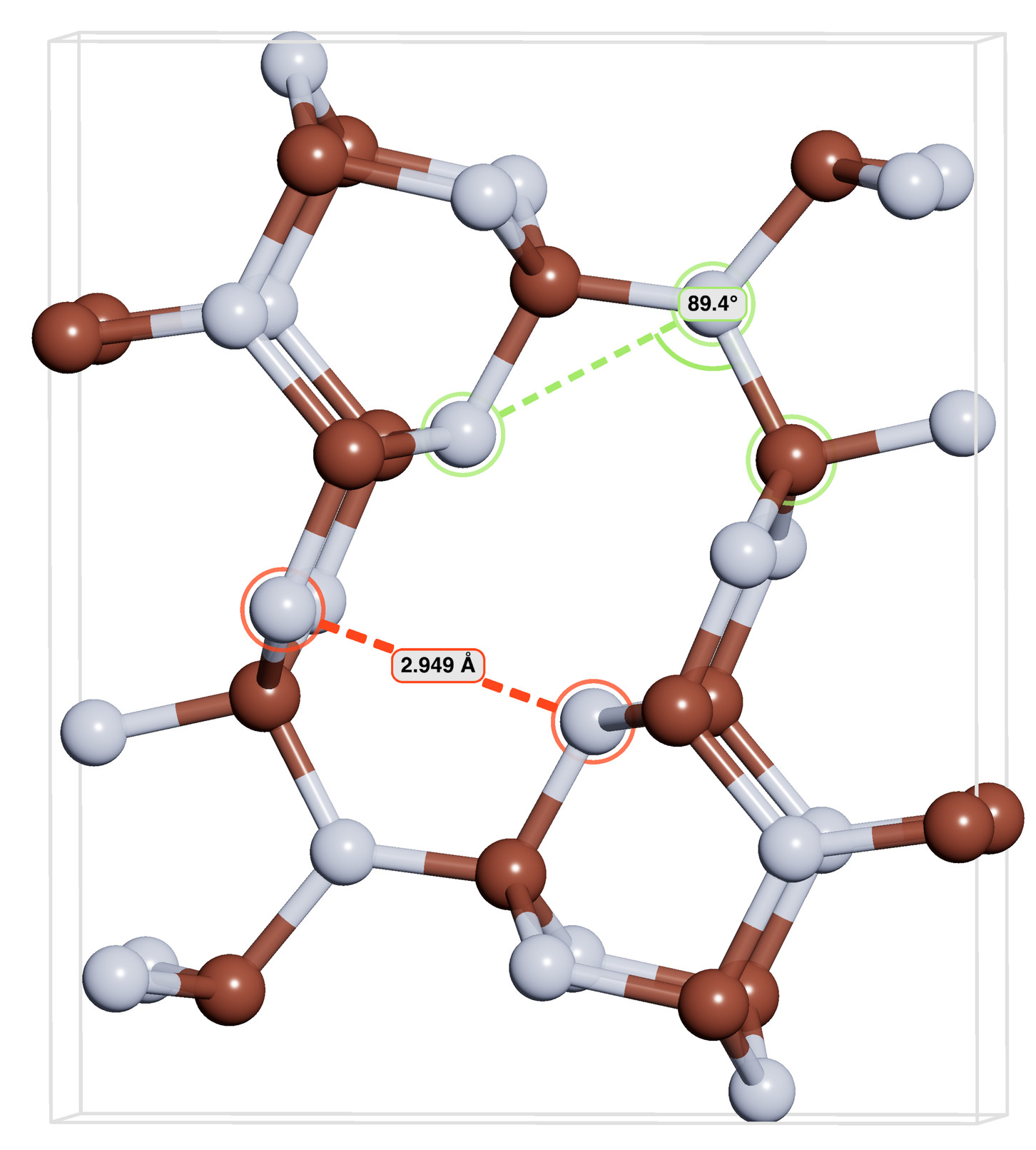}

    \caption{\textbf{Measurements:} Exported images with two different styles for the measurements. The user can select the colors, line style, highlighting type around the atoms, opacity and label size.}
    \label{fig:Measurements2}
\end{figure*}

\begin{figure*}
    \centering
    \includegraphics[width=1\linewidth]{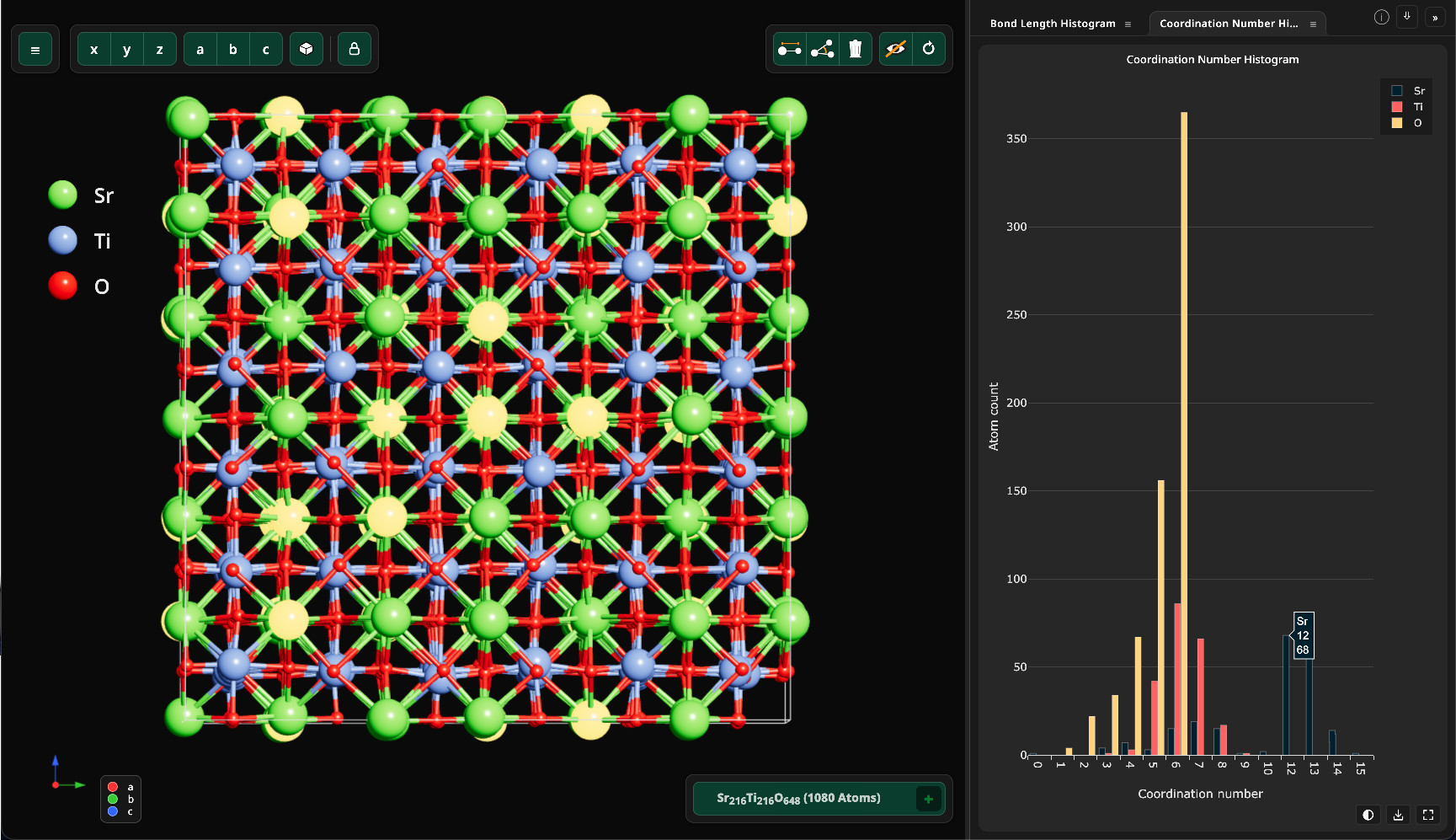}
    \includegraphics[width=1\linewidth]{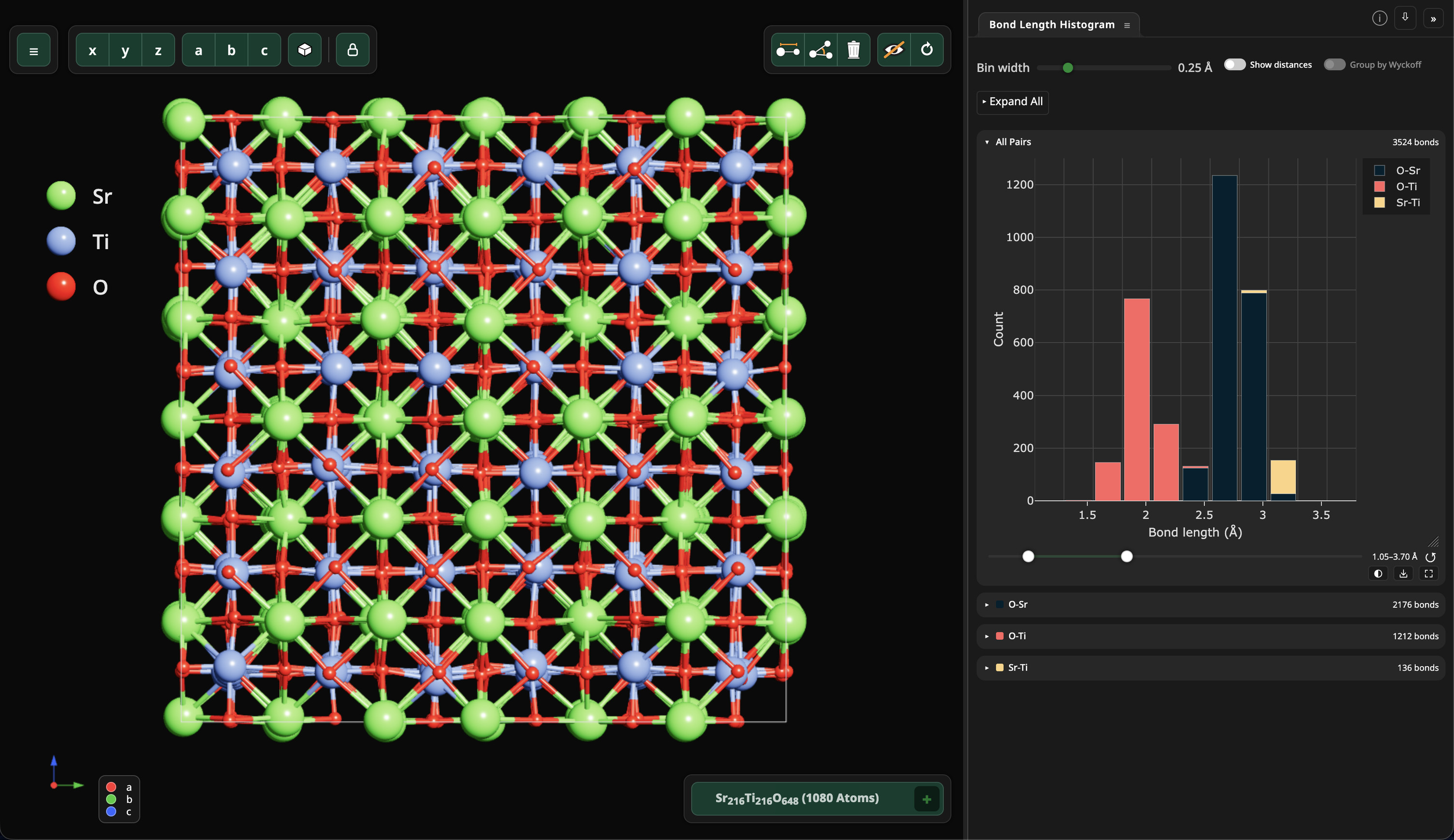}
    \caption{\textbf{Bond Lengths and Coordination Numbers:} \crysviz allows the analysis of coordination numbers(top) and bond lengths (middle) via histogram. Clicking a histogram bar highlights the atoms in this category in the structure view. The example shows a rattled supercells of SrTiO$_3$.}
    \label{fig:bond_length_coordination}
\end{figure*}

\begin{figure*}
    \centering
    \includegraphics[width=1\linewidth]{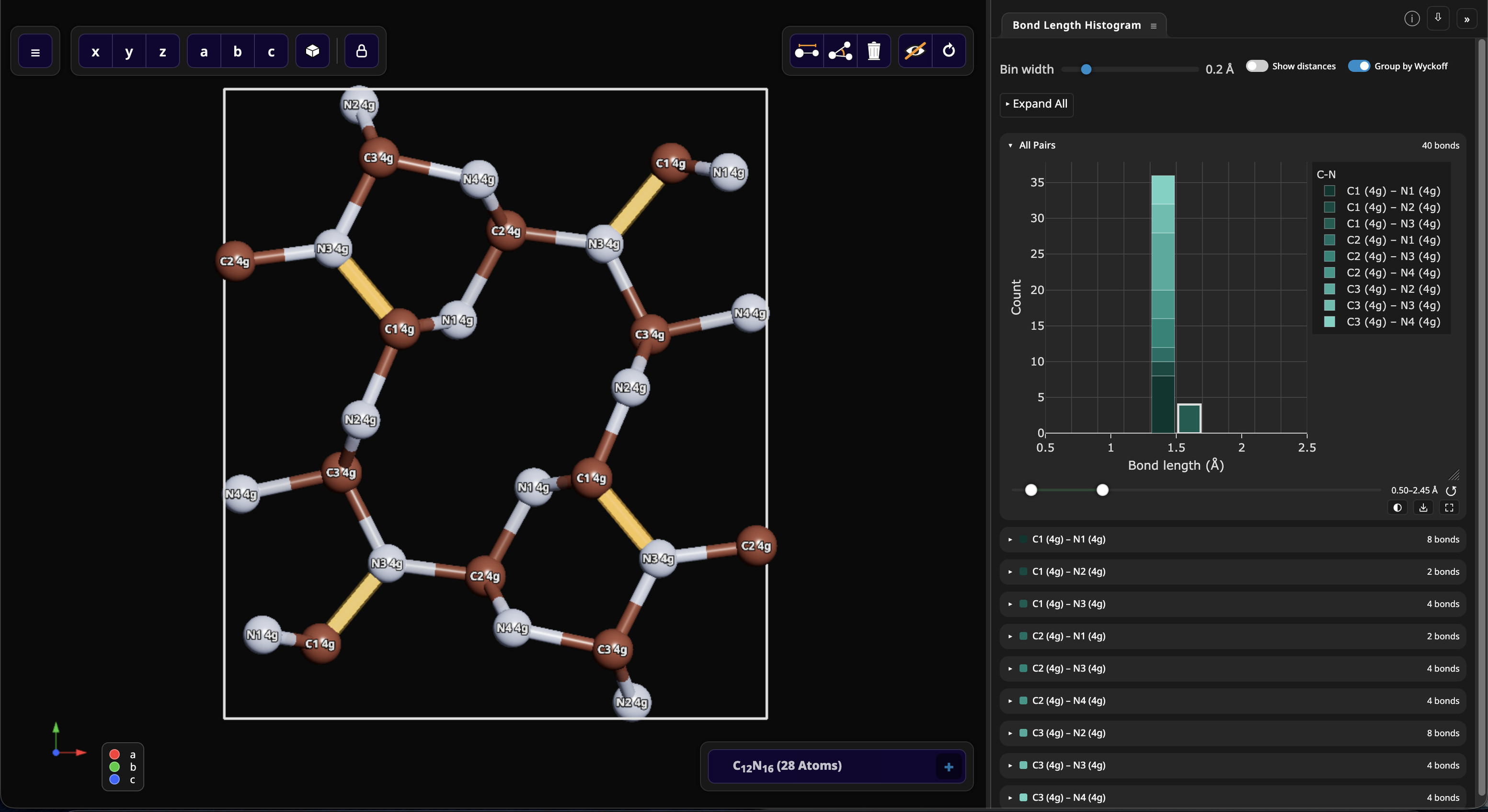}
    
    \caption{\textbf{Bond Lengths and Coordination Numbers:} When the Wyckoff Mode is active, atoms can be grouped by Wyckoff positions as shown here for the 4g Wyckoff positions in \textit{oP}28-C$_3$N$_4$. The selected bar corresponds to the highlighted bonds in the structure view.}
    \label{fig:bond_length_coordination_wyk}
\end{figure*}

\begin{figure*}
    \centering
    \includegraphics[width=1\linewidth]{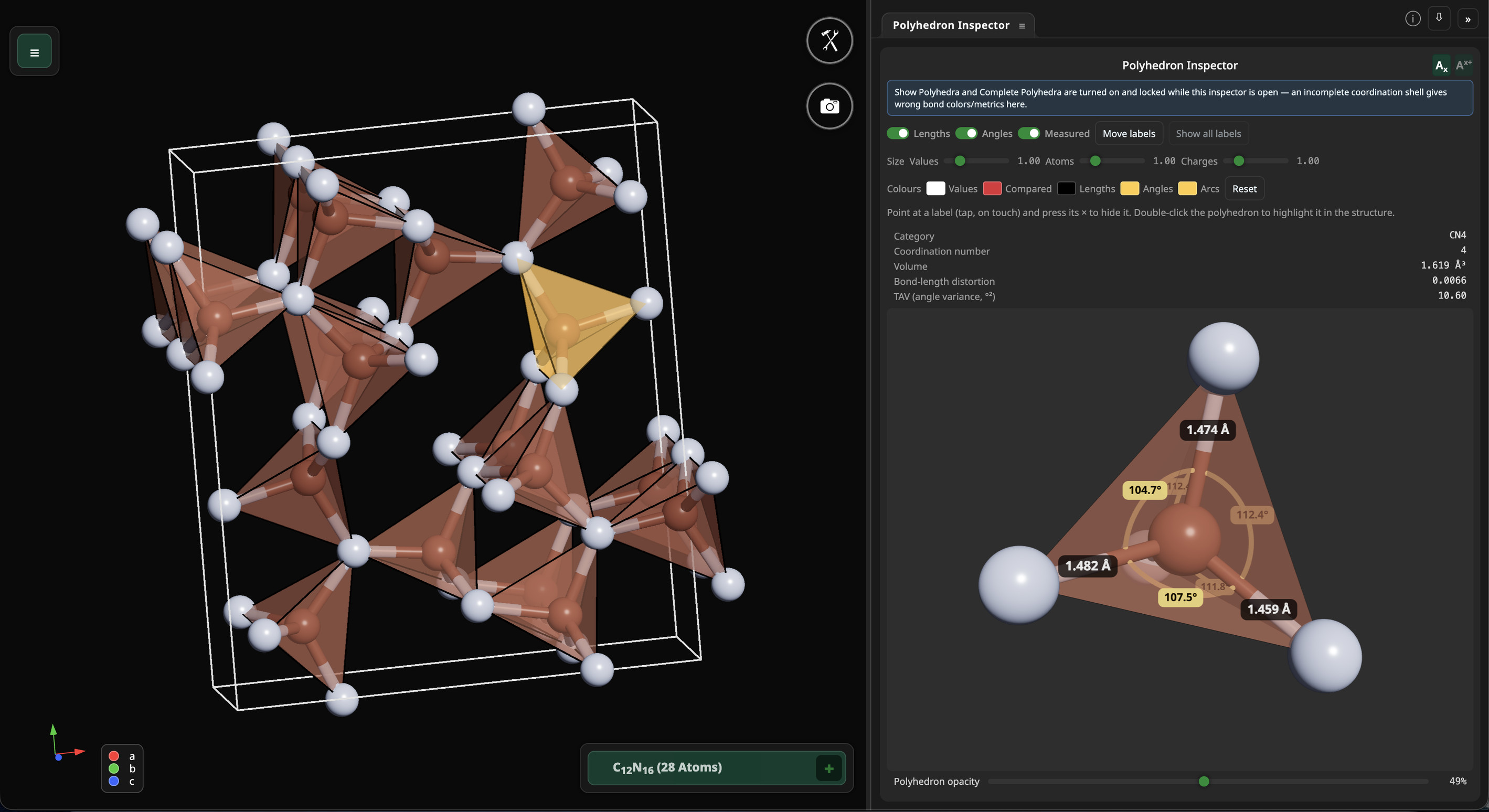}
    \caption{\textbf{Polyhedra Inspector:} Beyond the building block inspector, \crysviz features a polyhedra inspector that allows the detailed visualisation of a selected polyhedron including all measurements. For tetra- and octahedra bond length distortion and tetrahedral angle variance (TAV is calculated). The example shows one CN$_4$ tetrahedra in \textit{oP}28-C$_3$N$_4$.}
    \label{fig:tetrahedra_inspector}
\end{figure*}

\begin{figure*}
    \centering
\includegraphics[width=1\linewidth]{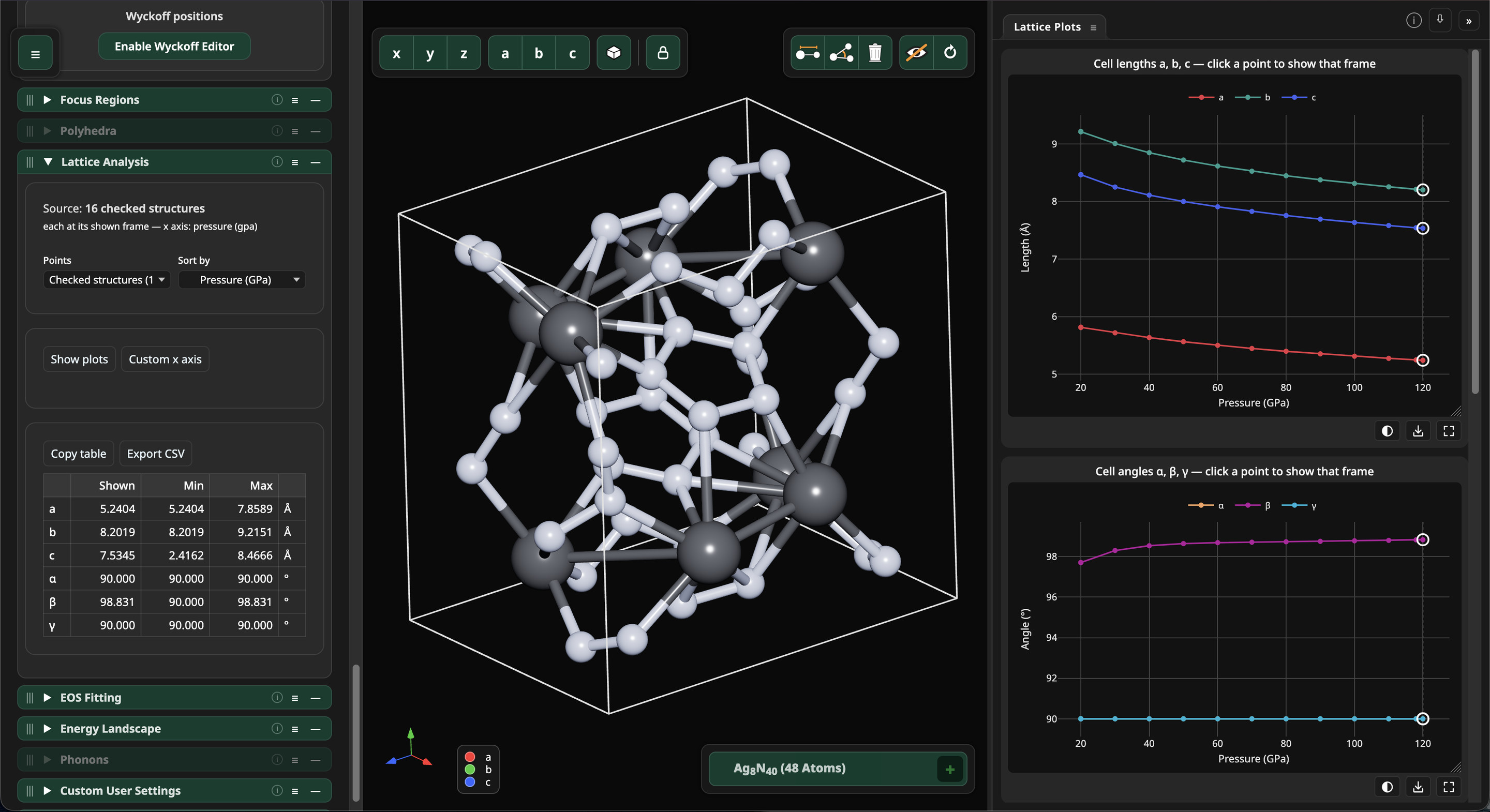}
    \caption{\textbf{Lattice Analysis:} Lattice constants and angles of several loaded structures, e.g., from the variable-cell relaxations of an equation of state calculation, plotted as a function of pressure, volume, file order, and many more. The example shows the calculated pressure-dependence of the lattice constants in AgN$_5$~\citep{liang2026high} between 20 and 120 GPa.}
    \label{fig:lattice}
\end{figure*}

\begin{figure*}
    \centering
    \includegraphics[width=1\linewidth]{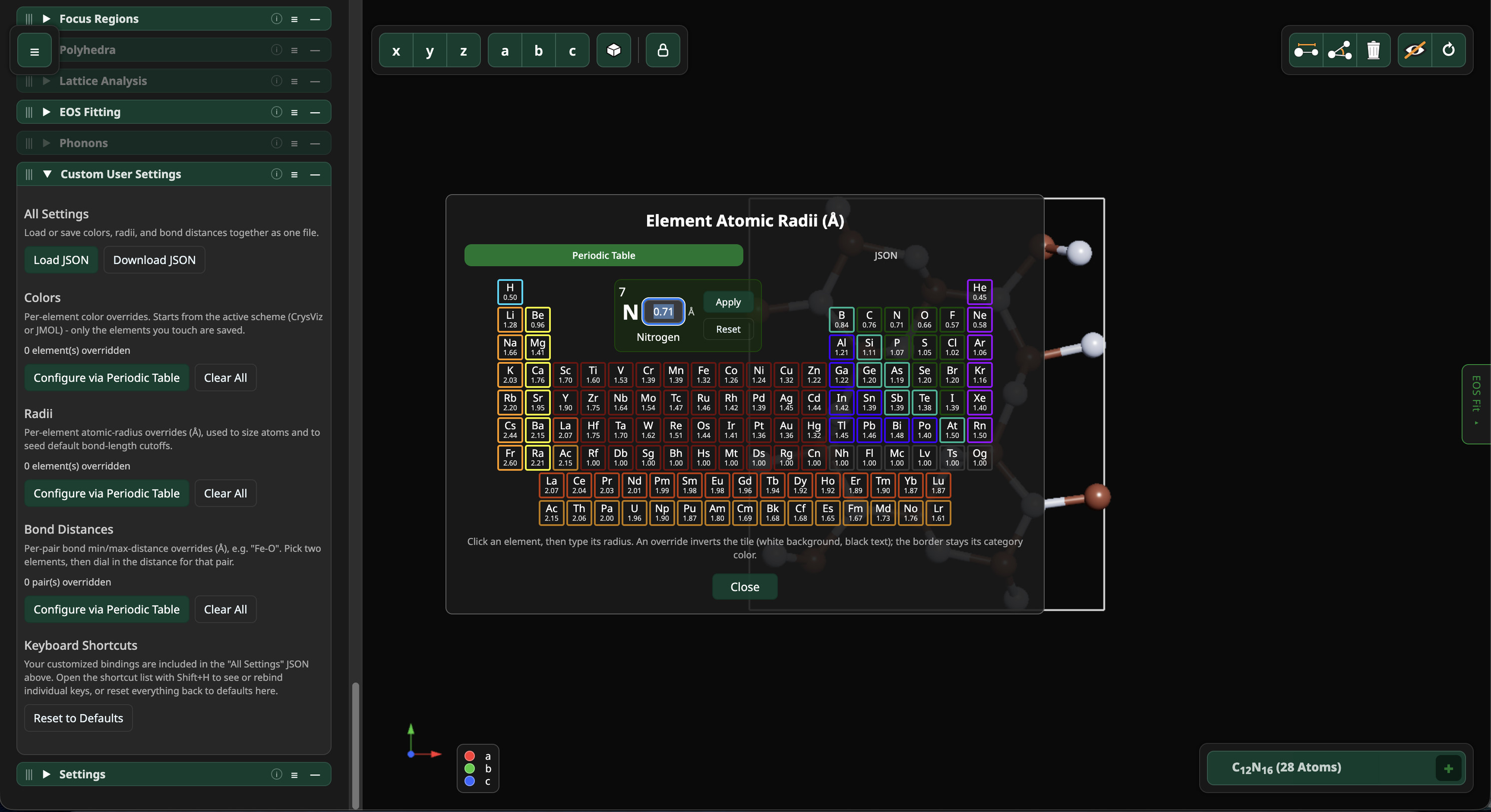}
    \caption{\textbf{User Default Settings:} The users can choose custom settings for color, bond cutoffs, short cuts, and atomic radii. The example shows the periodic table style selector for the atomic radius selection.}
    \label{fig:user-settings}
\end{figure*}

\begin{figure*}
    \centering
    \includegraphics[width=1\linewidth]{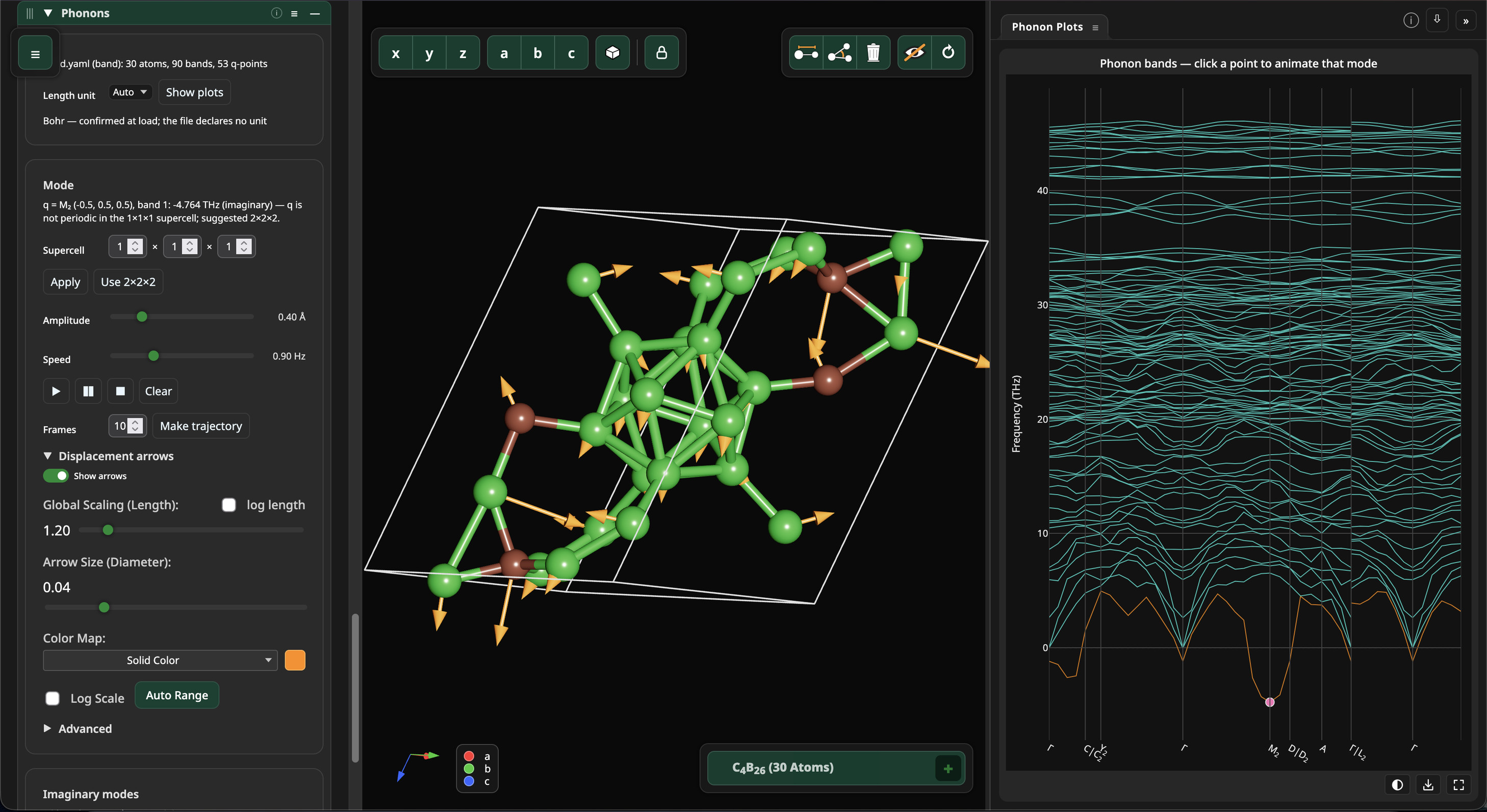}
    \caption{\textbf{Phonons Visualisation:} There is a \crysviz module that allows loading the \texttt{band.yaml} file from a phonopy calculation to plot the dispersion relation and visualise the atomic displacements related to a particular mode at a chosen k-point. The example shows the dispersion relation of C$_2$B$_{13}$ from \citet{liang2026BC}.}
    \label{fig:phonons}
\end{figure*}

\begin{figure*}
    \centering
    \includegraphics[width=1\linewidth]{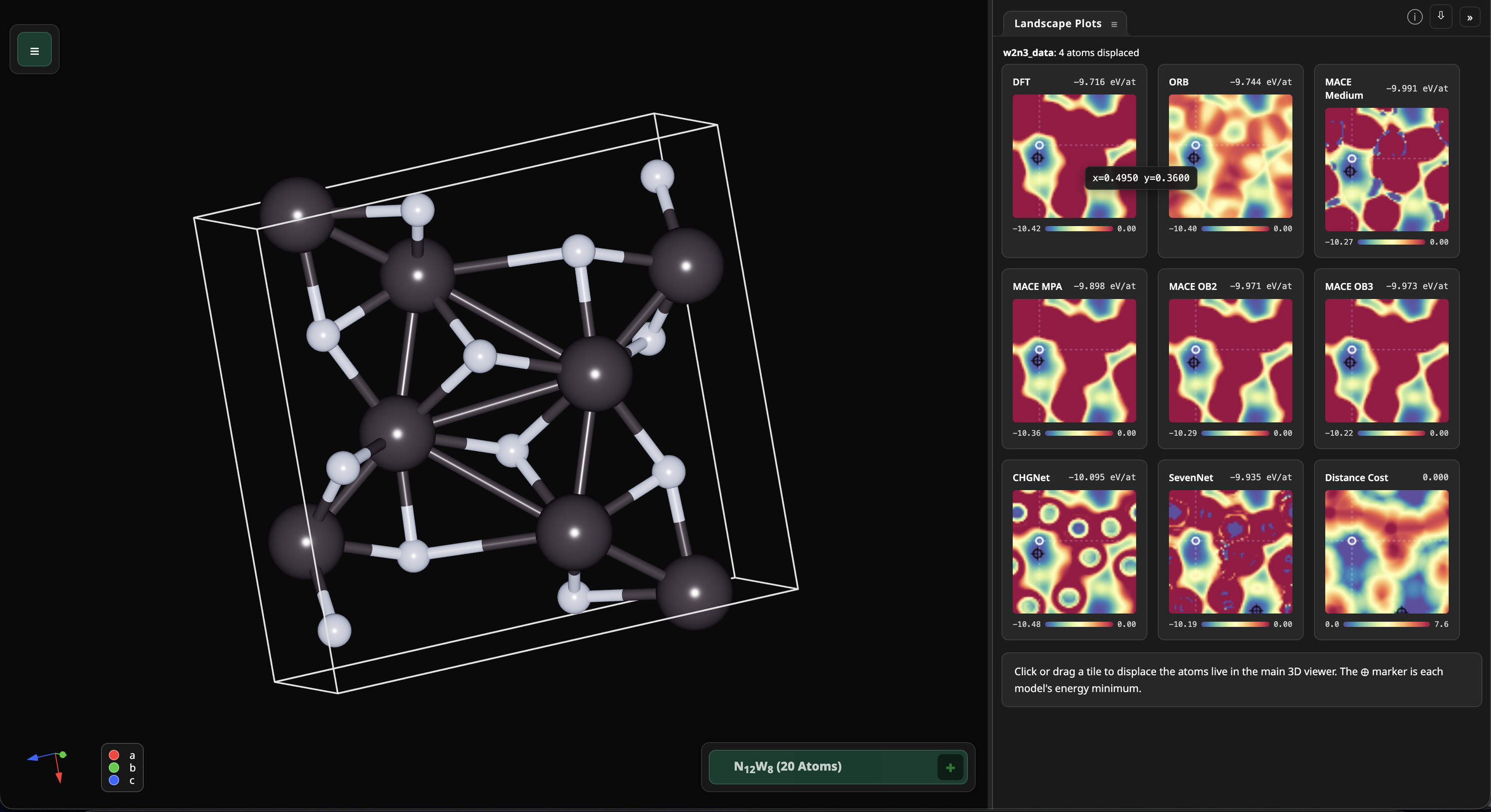}
    \caption{\textbf{Energy Landscapes:} There is a \crysviz module to visualise symmetry-restricted energy landscapes. This is a specific application for another project of a subset of the authors \citep{parackal2026symmetry}. It highlights how the modular system allows to expand \crysviz to support additional special needs or specific users. The example shows different symmetry-restricted energy landscapes for W$_2$N$_3$ calculated with DFT and uMLIPs.}
    \label{fig:energy_landscapes}
\end{figure*}

\begin{figure*}
    \centering
    \includegraphics[width=1\linewidth]{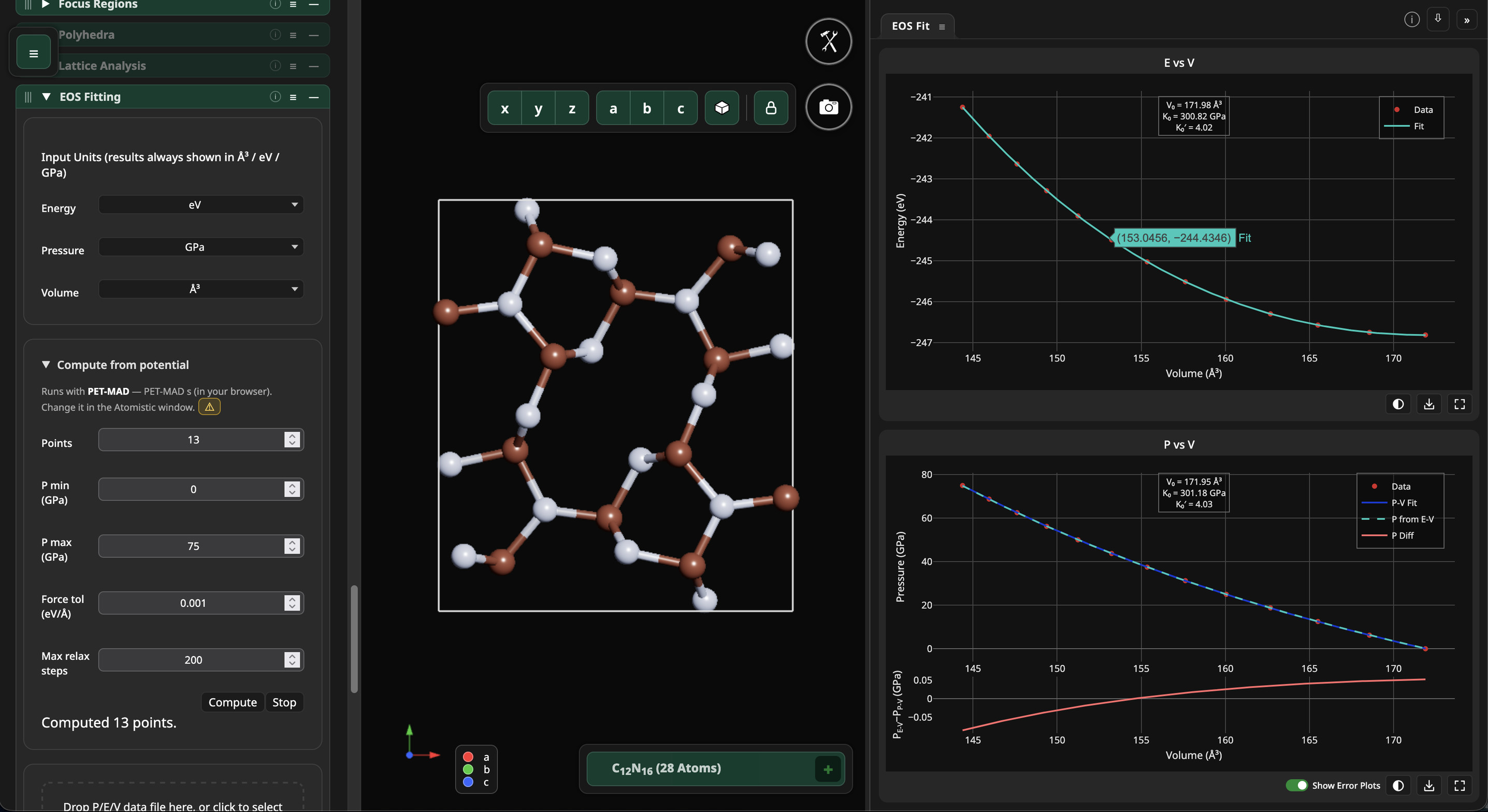}
    \caption{\textbf{Equation of state:} There is a \crysviz module to analyse and perform equation of state calculations. The calculations are performed with the currently selected uMLIP, here PET-MAD S is used, and a third order Birch-Murnaghan EOS is fitted to the data for \textit{oP}28-C$_3$N$_4$. The potential underestimates the bulk modulus by about 12\%.}
    \label{fig:EOS}
\end{figure*}

\end{document}